\documentclass[fleqn,usenatbib]{mnras}

\usepackage[T1]{fontenc}
\usepackage{ae,aecompl}

\usepackage{savesym} 
\usepackage{graphicx}	
\usepackage{amsmath}	
\usepackage{amssymb}	
\usepackage{amstext}
\usepackage{longtable}
\usepackage{multirow} 

\newcommand{\sn}{SN 2016X}
\newcommand{\host}{UGC 08041}
\newcommand\arcdeg{\mbox{$^\circ$}}
\newcommand{\ha} {\mbox{H$\alpha$}}
\newcommand{\hb} {\mbox{H$\beta$}}

\newcommand{\ld}{\mbox{$\lambda$}}
\newcommand{\eg}{{\textrm e.g.}}
\newcommand{\ie}{{\textrm i.e.}}
\newcommand{\msun}{\mbox{$M_{\odot}$}}
\newcommand{\rsun}{\mbox{$R_{\odot}$}}
\newcommand{\kms}{\mbox{$\rm{\,km\,s^{-1}}$}}
\newcommand{\ergs}{\mbox{$\rm{\,erg\,s^{-1}}$}}

\newcommand{\nickel}{\mbox{$^{56}$Ni}}
\newcommand{\cobalt}{\mbox{$^{56}$Co}}
\newcommand{\iron}{\mbox{$^{56}$Fe}}
\newcommand{\nodata}{ ~$\cdots$~ }

\title[SN 2016X in UGC 08041]{SN 2016X: A Type II-P Supernova with A Signature of Shock Breakout from Explosion of A Massive Red Supergiant}

\author[Huang et al.]{
F. Huang,$^{1,2,3}$\thanks{E-mail: huangfang@mail.tsinghua.edu.cn}
X.-F. Wang,$^{1}$\thanks{E-mail: wang\_xf@mail.tsinghua.edu.cn}
G. Hosseinzadeh,$^{4,5}$ P. J. Brown,$^{6}$ J. Mo,$^{1}$ J.-J. Zhang,$^{3,7,8}$
\newauthor K.-C. Zhang,$^{1}$ T.-M. Zhang,$^{9}$ D.-A. Howell,$^{4,5}$ I. Arcavi,$^{4,5,10}$ C. McCully,$^{4,5}$ S. Valenti,$^{11}$ 
\newauthor L.-M. Rui,$^{1}$ H. Song,$^{1}$ D.-F. Xiang,$^{1}$ W.-X. Li,$^{1}$ H. Lin,$^{1}$ L.-F. Wang$^{6}$
\\
$^{1}$Physics Department and Tsinghua Centre for Astrophysics, Tsinghua University, Beijing, 100084, China\\
$^{2}$Department of Astronomy, Shanghai Jiao Tong University, Shanghai, 200240, China \\
$^{3}$Key Laboratory for the Structure and Evolution of Celestial Objects, Chinese Academy of Sciences, Kunming 650011, China\\
$^{4}$Department of Physics, University of California, Santa Barbara, CA 93106-9530, USA\\
$^{5}$Las Cumbres Observatory, 6740 Cortona Dr., Suite 102, Goleta, CA 93117-5575, USA\\
$^{6}$George P. and Cynthia Woods Mitchell Institute for Fundamental Physics \& Astronomy, Texas A\&M University, Department of Physics and Astronomy, 4242 TAMU, College Station, TX 77843, USA\\
$^{7}$Yunnan Observatories, Chinese Academy of Sciences, Kunming 650011, China\\
$^{8}$Centre for Astronomical Mega-Science, Chinese Academy of Sciences, 20A Datun Road, Chaoyang District, Beijing, 100012, China\\
$^{9}$Key Laboratory of Optical Astronomy, National Astronomical Observatories, Chinese Academy of Sciences, Beijing 100012, China \\
$^{10}$Einstein Fellow \\
$^{11}$Department of Physics, University of California, 1 Shields Avenue, Davis, CA 95616-5270, USA \\
}

\date{Accepted XXX. Received YYY; in original form ZZZ}

\pubyear{2018}

\begin{document}
\label{firstpage}
\pagerange{\pageref{firstpage}--\pageref{lastpage}}
\maketitle

\begin{abstract}
We present extensive ultraviolet (UV) and optical photometry, as well as dense optical spectroscopy for type II Plateau (IIP) supernova \sn\ that exploded in the nearby ($\sim$ 15 Mpc)  spiral galaxy \host. The observations span the period from 2 to 180 days after the explosion; in particular, the \emph{Swift} UV data probably captured the signature of shock breakout associated with the explosion of \sn. It shows very strong UV emission during the first week after explosion, with contribution of $\sim$ 20 -- 30\% to the bolometric luminosity (versus $\lesssim$ 15\% for normal SNe IIP). Moreover, we found that this supernova has an unusually long rise time of about 12.6 $\pm$ 0.5 days in the $R$ band (versus $\sim$ 7.0 days for typical SNe IIP). The optical light curves and spectral evolution are quite similar to the fast-declining type IIP object SN 2013ej, except that \sn\ has a relatively brighter tail. Based on the evolution of photospheric temperature as inferred from the $Swift$ data in the early phase, we derive that the progenitor of \sn\ has a radius of about 930 $\pm$ 70 \rsun. This large-size star is expected to be a red supergiant star with an initial mass of $\gtrsim$ 19 -- 20 M$_{\odot}$ based on the mass $--$ radius relation of the Galactic red supergiants, and it represents one of the most largest and massive progenitors found for SNe IIP.
\end{abstract}



\section{Introduction}

Type II supernovae (SNe) are the outcome of massive stars (with initial mass $\geq$ 8 \msun; \eg\  \citealt{1984ApJ...277..791N, 1988PhR...163...13N, 2009ARA&A..47...63S, 2013ApJ...765L..43I}) experiencing gravitational core collapse after energy exhaustion at the end of life. They are characterized by P-cygni profile of Balmer lines in the early optical spectra compared to type I SNe \citep{1997ARA&A..35..309F}. Based on the behaviors of light curves, SNe II are further divided into two subclasses: those with a prolonged plateau lasting $\sim$ 100 days are called type IIP, while those with a linear decline trend after maximum belong to type IIL \citep{1979A&A....72..287B}. Recently, statistical work with large samples from different surveys tends to favour for a continuum distribution of the observational properties of SNe II (\eg\ \citealt{2014ApJ...786...67A, 2015ApJ...799..208S, 2016MNRAS.459.3939V}). As the most abundant sub-type, SNe IIP occupy about 70\% of all observed SNe II in a volume-limited sample \citep{2011MNRAS.412.1441L}. The observed plateau in the light curve results from the propagation of a cooling and recombination wave through the SN envelope \citep{1971Ap&SS..10...28G, 1976Ap&SS..44..409G}. The presence of prominent hydrogen lines indicates that they retain a significant fraction of hydrogen envelopes before explosion.

Analysis of the archive images allow direct detections of the progenitors for a few SNe IIP, which are generally found to be red supergiant (RSG) stars with a mass range of 8.5--16.5 \msun \citep{2009ARA&A..47...63S, 2015PASA...32...16S}. The observational limit is lower than the prediction from theoretical models, e.g., 8 to 25 \msun \citep{2012A&A...537A.146E}. This inconsistency might be somewhat related to the presence of substantial circumstellar dust around the RSGs, which could lead to the underestimate of luminosity and hence the initial mass of the progenitor stars \citep{2012ApJ...759L..13F, 2012ApJ...756..131V, 2014ApJ...787..139D}. SNe IIP show a large diversity in the observational properties, such as peak luminosity, plateau length, expansion velocity, and synthesized nickel mass \citep{2003ApJ...582..905H}. These are connected with the explosion mechanism and the physical characteristics of the progenitors such as mass, explosion energy, and initial radius \citep{2009ApJ...703.2205K, 2011ApJ...741...41P}. Dozens of SNe IIP have been extensively studied from the ultraviolet to the near-infrared wavelength, \ie\ SN 2005cs \citep{2009MNRAS.394.2266P}, SN 2009N \citep{2014MNRAS.438..368T}, and SN 2013ej \citep{2014MNRAS.438L.101V, 2015ApJ...807...59H}, which helps take a deep look into the observed diversity and the progenitor physics. \sn\ provides another opportunity for such kind of study.

\sn\ (ASASSN-16at) was discovered by All Sky Automated Survey for SuperNovae (ASAS-SN) on 2016 Jan. 20.59 (UT dates are used throughout this paper) in the nearby SBd galaxy \host\ (z=0.004408 from NED) at a $V$-band magnitude of $\sim$15.1 mag. The J2000 coordinates of the SN are $\alpha$ = 12$^h$55$^m$15.50$^s$ and $\delta = +00\arcdeg 05 \arcmin 59.7 \arcsec$, approximately 60$\arcsec\,$ south and 42$\arcsec\,$ east from the centre of \host\ \citep{2016ATel.8566....1B}. The last non-detection was reported on Jan. 18.35 with a limit of $V > 18.0$ mag, but it was detected on 2016 Jan. 19.49 at $V$ $\sim$ 16.6 mag and Jan. 19.50 at $V \sim$ 17.0 mag. We therefore adopt 2016 Jan. 18.9 (MJD = 57405.92 $\pm$ 0.57) as the explosion time. An optical spectrum obtained on Jan. 20.75 suggests that it is a young core-collapse SN \citep{2016ATel.8567....1H}, while another spectrum obtained on Jan. 23.88 confirms that it is a type II-P SN \citep{2016ATel.8584....1Z}. \citet{2016ATel.8588....1G} reported the discovery of X-rays from \sn\ with \emph{Swift}, which indicates that \sn\ may have experienced moderate interaction with circumstellar material or stellar wind at early phase. We therefore triggered an instant follow-up campaign to study the photometric and spectroscopic evolution of this young type II-P supernova. The distance to its host galaxy is estimated to be 15.2 Mpc (distance modulus $\mu$ = 30.91 $\pm$ 0.43 mag) using Tully-Fisher method \citep{2014MNRAS.444..527S}, which is adopted throughout this work.

In this paper, we present photometry and spectroscopy of the nearby type IIP \sn. In Section \ref{sec:obs}, we describe the observations and data reduction process for photometric and spectroscopic data. In Section \ref{sec:lc}, we study the photometric behavior of \sn. The spectroscopic evolution is presented in Section \ref{sec:spec}. We discuss the explosion parameters and progenitor properties of \sn\  in Section \ref{sec:discuss}, and summarize our conclusions in Section \ref{sec:summ}.

\section{Observations and Data Reduction} \label{sec:obs}

\subsection{Photometry} \label{subsec:phot}
\subsubsection{Ground-based Observation} \label{subsubsec:optobs}
High-cadence, broad-band photometric data of \sn\ was obtained in Johnson $UBV$ and Sloan $gri$ filters with the 1.0~m telescopes of Las Cumbres Observatory (LCO; \citealt{2013PASP..125.1031B}), spanning from 2016 Jan. 21 to 2016 Jul. 6. We also used the 0.8~m Tsinghua University-NAOC telescope (TNT; \citealt{2008ApJ...675..626W, 2012RAA....12.1585H}) at Xinglong Observatory and the Lijiang 2.4~m telescope (LJT; \citealt{2015RAA....15..918F}) of Yunnan Astronomical Observatories in China to collect photometry in Johnson-Cousin $UBVRI$ filters. The observations began on 2016 Jan. 23 and ended on 2016 Jun. 3.

\begin{figure*}
\centering
\includegraphics[width=0.5\textwidth,trim=500 0 500 0]{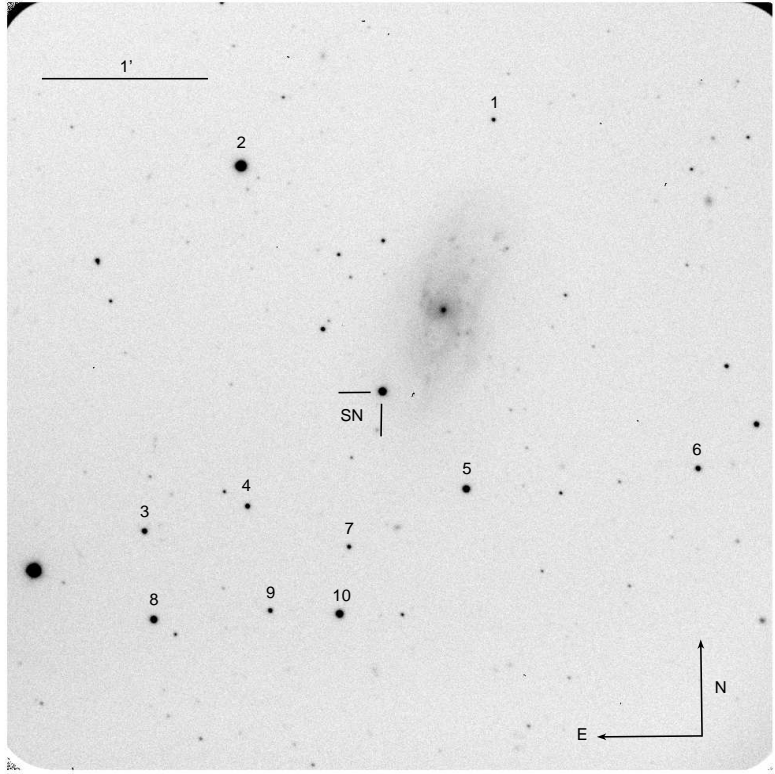}
\caption{\sn\ in \host. The $R$-band image of the field of SN 2016X was taken on 2016 Mar. 3 with the Lijiang 2.4~m Telescope. \sn\ and the 10 reference stars are marked.} \label{fig:finder}
\end{figure*}

All data were pre-processed with standard \textsc{IRAF}\footnote{IRAF is distributed by the National Optical Astronomy Observatories, which are operated by the Association of Universities for Research in Astronomy, Inc., under cooperative agreement with the National Science Foundation (NSF).} routines, including the corrections for bias, overscan, flat-field, and cosmic-ray removal. For TNT and LJT data, instrumental magnitudes were determined using the point-spread function (PSF) photometry with the \textsc{SNOoPy} package\footnote{http://sngroup.oapd.inaf.it/snoopy.html}. The LCO data were reduced using \texttt{lcogtsnpipe} \citep{2016MNRAS.459.3939V}. The colour terms and extinction coefficients were derived from observations of Landolt stars on photometric nights \citep{1992AJ....104..340L}. The photometric zeropoints were determined by comparing the magnitudes of 10 field stars (marked in Figure \ref{fig:finder}) to the values transformed from the Sloan Digital Sky Survey (SDSS) Data Release 9 catalogue \citep{2012ApJS..203...21A} using the relation from \citet{2008AJ....135..264C}. The coordinates and magnitudes of the reference stars around \sn\ are listed in Table \ref{tab:standstar}, and the final calibrated magnitudes of \sn\ are presented in Table \ref{tab:phot_tnt}--\ref{tab:phot_LCO}.

\subsubsection{Swift UVOT Observations} \label{subsubsec:uvobs}
\sn\ was also observed in the ultraviolet and optical bands with the Ultra-Violet/Optical Telescope (UVOT; \citealp{2005SSRv..120...95R}) on board the \emph{Swift} spacecraft \citep{2004ApJ...611.1005G}. The space-based observations were obtained in the $uvw2, uvm2, uvw1, u, b$, and $v$ filters, covering the period from 2016 Jan. 21 to 2016 Mar. 5, and these data were taken from the \emph{Swift} Optical/Ultraviolet Supernova Archive\footnote{http://swift.gsfc.nasa.gov/docs/swift/sne/swift\_sn.html} (SOUSA; \citealp{2014Ap&SS.354...89B}). The data reduction is based on the method described in \citet{2009AJ....137.4517B}, including subtraction of the host galaxy count rates and usage of the revised UV zeropoints and time-dependent sensitivity loss from \citet{2011AIPC.1358..373B}. The UVOT magnitudes of \sn\ are listed in Table \ref{tab:uvot}.

\subsection{Spectroscopy} \label{subsec:spec}
The spectroscopic observations of \sn\ started on 2016 Jan. 20 and continued until 2016 Jun. 9, corresponding to $\sim$2 day to $\sim$140 days after the explosion. A total of 40 low-resolution optical spectra were collected using the LCO 2~m Faulkes Telescope North (FTN; with FLOYDS), the Lijiang 2.4~m telescope (with YFOSC; \citealp{2016PASP..128k5005F}), and the Xinglong 2.16~m telescope (with BFOSC). A journal of spectroscopic observations is given in Table \ref{tab:spelog}.

The spectroscopic data were reduced in a standard manner under the \textsc{IRAF} environment. After bias and overscan corrections, flat-fielding and cosmic-ray removal, one dimensional spectra were extracted using the optimal extraction method \citep{1986PASP...98..609H}. The wavelength calibration was done using the Fe/Ar and Hg/Ar lamp spectra, and the fluxes were calibrated using spectrophotometric standards observed on the same night with the same instrumental set-up. FLOYDS spectra were reduced using the \texttt{floydsspec} pipeline.

\section{Photometric Evolution}  \label{sec:lc}
\begin{figure}
\centering
\includegraphics[width=0.5\textwidth]{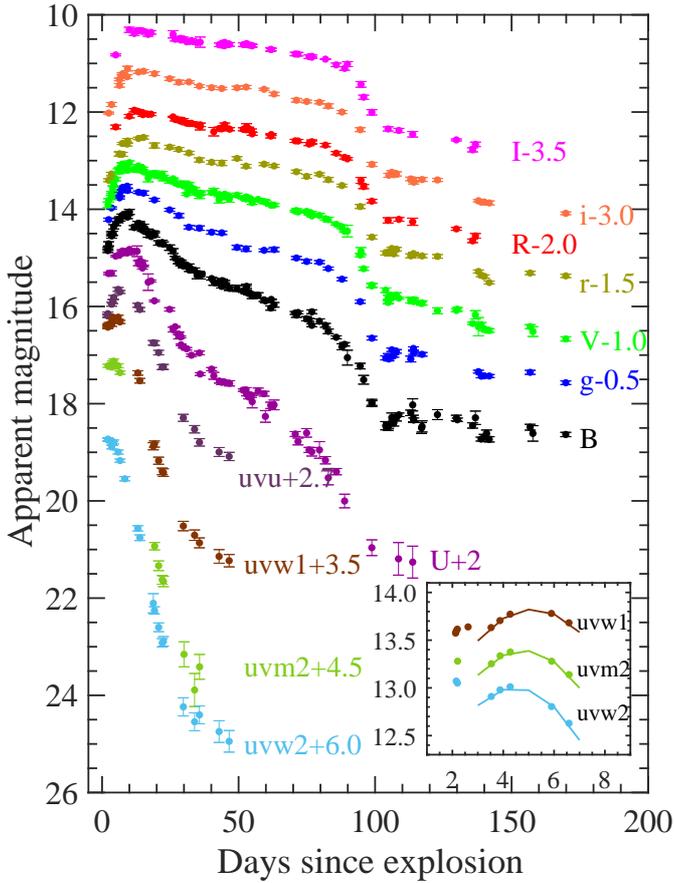}
\caption{Light curves of \sn\ in ultraviolet (UV) and optical bands. The insert is a zoom on the early-time UV light curve, with polynomial fitting to the data around maximum. Prominent UV emission is clearly seen at a few days before the primary UV peaks. The phase is given relative to the estimated explosion date, MJD = 57,405.92.}  \label{fig:lc}
\end{figure}

The light curves of \sn\ in UV and optical bands are shown in Figure \ref{fig:lc}, ranging from 2 to 170 days after explosion. The UV luminosity rises to the peak in a short time, followed by a rapid decline. The optical light curves resemble the evolution of typical SNe IIP but with relatively faster declines during the plateau phase. We present detailed analysis in the following subsections.

\subsection{Swift UV Light Curves }\label{subsec:uvlc}
The \emph{Swift} UVOT observations of \sn\ were triggered immediately after its discovery. The very early light curves in the $uvw2$ and $uvm2$ bands show an initial decline before rising to the peak at t $\sim$ 5 days after explosion (see the insert panel of Figure \ref{fig:lc}). This indicates that \sn\ may have another UV peak within 1--2 days from the explosion, which could be due to the breakout of a blast shockwave through the progenitor star's outer envelope after the core-collapse explosion \citep{1977ApJS...33..515F, 1978ApJ...223L.109K}. The observed UV trough might thus be associated with the cooling of shock breakout, when the temperature behind the shock is lower than that at the shock front \citep{2008Sci...321..223S}. Such a UV trough had ever been reported for two SNe IIP at relatively larger distances, \ie\ SNLS-04D2dc \citep{2008Sci...321..223S} and SNLS-06D1jd \citep{2008ApJ...683L.131G}.

By adopting a polynomial fitting to the early data, we obtained m$_{uvw2}$(max) = 12.79 mag on 4.53 d, m$_{uvm2}$(max) = 12.61 mag on 4.88 d, and m$_{uvw1}$(max) = 12.68 mag on 5.02 d relative to the explosion date. The rise time for the primary UV peaks is $\sim$ 2 days longer than that of SNLS-04D2dc and SNLS-06D1jd, indicating that \sn\ may have a progenitor with a larger initial radius. After the maximum, the SN declines quickly in the \emph{swift} $uvw2, uvm2, uvw1$, and $u$ bands, with a rate of 0.245 $\pm$ 0.012, 0.269 $\pm$ 0.022, 0.208 $\pm$ 0.027, 0.135 $\pm$ 0.011 mag d$^{-1}$, respectively. While the corresponding decay rate is 0.047 $\pm$ 0.011, 0.016 $\pm$ 0.014 mag d$^{-1}$ in \emph{swift} $b$ and $v$ bands. Note that \sn\ shows a faster decline in $uvm2$ than in $uvw2$ and $uvw1$, which is against the usual trend that the decay rate steepens at shorter wavelengths. This opposite trend is also seen in other SNe IIP (\ie\ SN 2005cs), and it might be related to the fact that more Fe III and Fe II lines are concentrated within the $uvm2$ bandpass \citep{2007ApJ...659.1488B}.

\begin{figure}
\centering
\includegraphics*[width=0.5\textwidth]{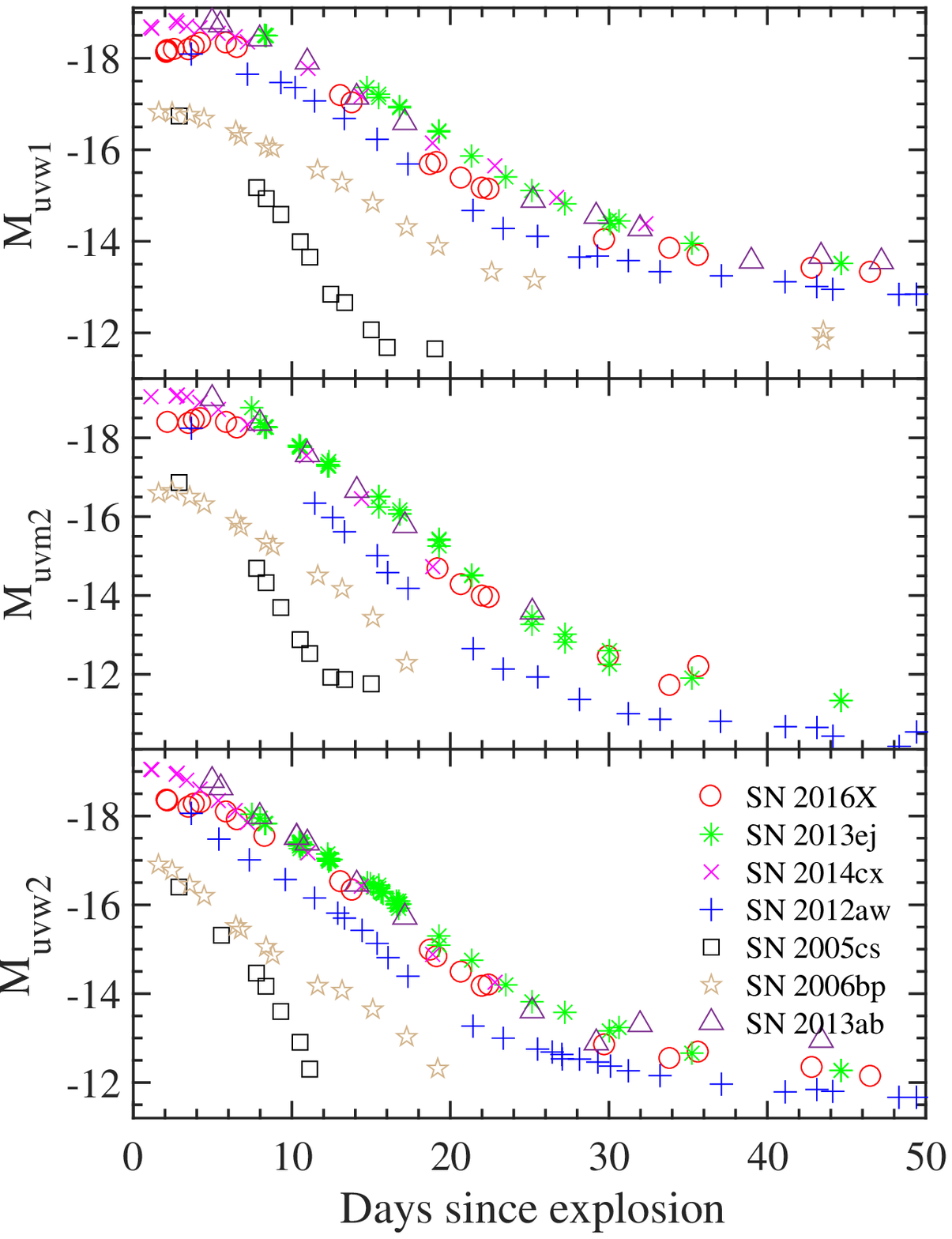}
\caption{Comparison of the UV absolute light curves of \sn\ with a few well-observed SNe IIP. For the comparison SNe, the distance modulus $\mu$ in mag, and extinction $A_V$ in mag are listed in the brackets after the name of each SN sample, and the references are: SN 2005cs (29.26, 0.095), \citet{2007ApJ...659.1488B}, \citet{2009MNRAS.394.2266P};SN 2006bp (31.22, 0.08), \citet{2007ApJ...664..435I};SN 2012aw (29.98, 0.08), \citet{2013ApJ...764L..13B}; SN 2013ab (31.90, 0.14), \citet{2015MNRAS.450.2373B};SN 2013ej (29.91, 0.19), \citet{2015ApJ...807...59H}; SN 2014cx (31.74, 0.31), \citet{2016ApJ...832..139H}.} \label{fig:absuv}
\end{figure}

Figure \ref{fig:absuv} shows \emph{Swift} UVOT absolute light curves of \sn\ and some well-observed SNe IIP. Extinction corrections have been applied to all of our objects. As it can be seen, \sn\ lies on the bright side of SNe IIP, and it reached the UV maximum 2--3 days later than other objects with UV observations. After t $\approx$ 1 month from the peak, the UV light curves seem to flatten out especially in the uvw1 and uvw2 filters, and this is similarly seen in SN 2012aw, SN 2013ab, and SN 2013ej. At this phase, the UV emission becomes very weak and the photometry can be significantly affected by optical photons leaked out of the red tails of the UV filters \citep{2016AJ....152..102B}.

\subsection{Optical Light Curves} \label{subsec:lcs}
The overall evolution of the optical light curves of \sn\ can be divided into four main phases: the rising phase ($\sim$ 15 days), the plateau phase ($\sim$ 90 days), the transitional phase ($\sim$ 100 days), and the nebular phase ($\ge$ 100 days).

The densely sampled data obtained immediately after the explosion allow us to catch the rising evolution of \sn\ in very early phase. Using polynomial fit to the observed data around the maximum light, we are able to estimate the dates of maximum light and the peak magnitudes in different filters. The results for the phases of maximum and peak magnitudes in different bands are listed in Table \ref{tab:photpara}.

After the maximum light, the $B$-band magnitude declines by $\sim$ 4.0 mag in 100 days, which is larger than the typical value for SNe IIP (\ie\ ${\beta}^B_{100} < 3.5$ mag; \citealt{1994A&A...282..731P}). The $V$-band declines by $\sim$ 0.8 mag from the peak brightness in the first 50 days after explosion, which is also larger than normal SNe IIP (i.e., $s$50$_V < 0.5$ mag; \citealt{2014MNRAS.445..554F}). Moreover, there are a few luminous SNe IIP (\eg\ SNe 2007od, 2007pk, 2009bw, 2009dd, and 2013ej) that are found to show similar large post-maximum magnitude declines \citep{2015MNRAS.448.2608V, 2015ApJ...807...59H}. This indicates that a larger V-band decline should be used to make a distinguish between SNe IIP and SNe IIL, or these fast-declining SNe IIP may actually represent a subclass linking normal SNe IIP and SNe IIL. From the end of the plateau phase, the SN starts a transitional phase with a very rapid flux drop. For example, the $V$-band magnitude drops by $\sim$ 2.0 mag during the phase from t$\approx$ +90 days to t$\approx$ +130 days. After t $\approx$ 110 days, the SN enters into the nebular phase powered by the radioactive decay (i.e.,\cobalt\ to \iron). The decline rates at this phase are estimated to be 0.79, 1.44, 1.22, and 1.14 mag (100d)$^{-1}$ in $BVRI$ bands, respectively.

\subsection{Rise time}
The rise time is an important parameter to constrain the properties of progenitor and explosion physics of SNe, which is typically defined as the time between the explosion epoch and the maximum light. Following the definition by \citet{2015A&A...582A...3G}, we adopt the maximum-light date as the time when the $r/R$-band magnitude rises by less than 0.01 mag per day.

Based on a sample of 20 SNe IIP and IIL, \citet{2015A&A...582A...3G} found that SNe II show a diversity of rise time, with an average value of 7.0 $\pm$ 0.3 days for SNe IIP. The rise time is found to depend more sensitively on the progenitor radius than the mass and explosion energy \citep{2011ApJ...728...63R}. On the other side, recent studies indicate that the rise time of SNe II only shows a weak correlation with their luminosities \citep{2016MNRAS.459.3939V, 2016ApJ...820...33R}. This is in contrast to previous conclusion that brighter SNe II tend to have longer rise time \citep{2011ApJ...736..159G, 2014MNRAS.438L.101V, 2015A&A...582A...3G}.

\begin{figure*}
\centering
\includegraphics[width=0.46\textwidth]{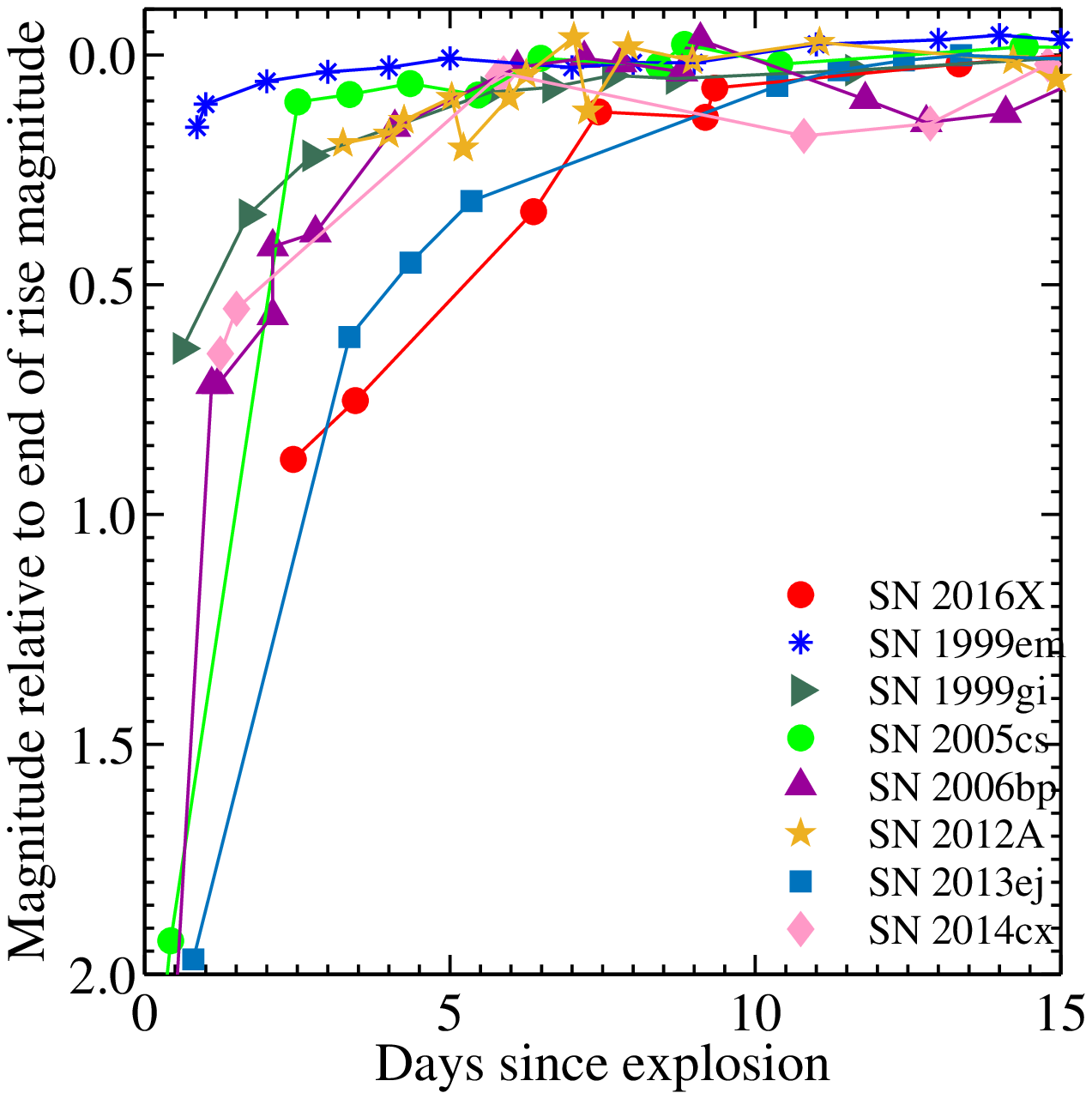}
\includegraphics[width=0.45\textwidth]{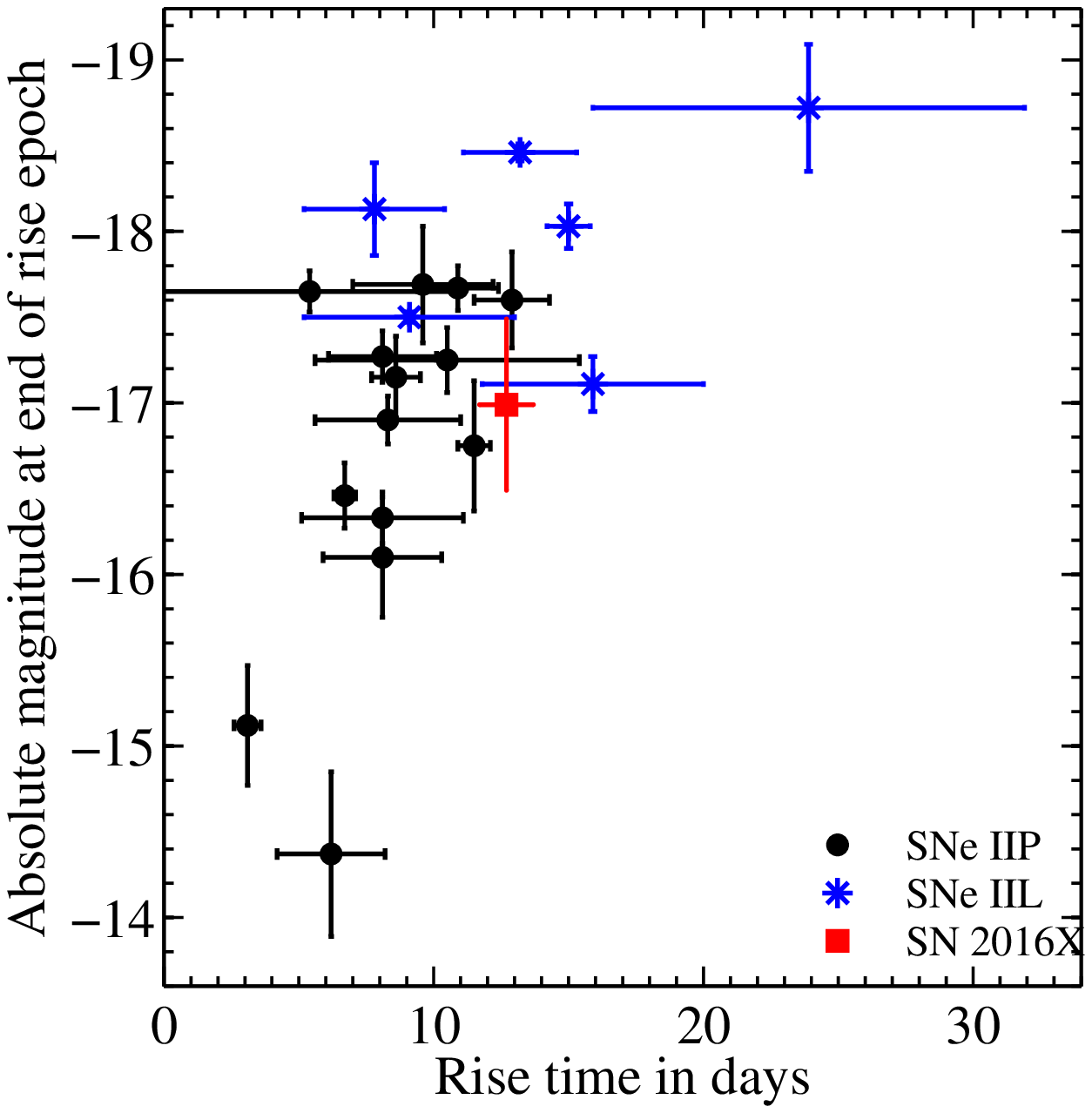}
\caption{Left: Comparison of the $r$/$R$-band light curve of \sn\ with a few SNe IIP. Right: Comparison of $r$/$R$-band absolute magnitudes at the end-of-rise epoch and rise time. The filled red square represents \sn\, while other dots represent a sample of SNe II from \citet{2015A&A...582A...3G}.}
\label{fig:trise}
\end{figure*}

Fitting a low-order polynomial to the data around maximum, we find that the $r$-band light curve has a rise time of 12.6 $\pm$ 0.5 days for \sn, and an absolute peak magnitude of $-$17.00 $\pm$ 0.43 mag. Figure \ref{fig:trise} shows the comparison of $r$/$R$-band light curves and rise time between \sn\ and some SNe II with early photometry. One can see that \sn\ has a longer rise time than typical SNe IIP, while the absolute magnitude at the end of rise follows the brighter-slower trend. The longer rise time of \sn\ indicates that its initial radius should be larger than that of normal SNe IIP, as predicted by the fact that photons take longer time to reach the surface of exploding star.

\subsection{Reddening and Colour Curves} \label{subsec:colo}
The Galactic reddening along the line of sight to \sn\ is $E(B-V)_{\rm MW}$ = 0.02 mag \citep{2011ApJ...737..103S}. The host galaxy reddening is estimated using the colour method raised by \citet{2010ApJ...715..833O} which assumes that the intrinsic $V-I$ colour is constant (\ie, $(V-I)_0$ = 0.656 mag) toward the end of the plateau phase. Fitting the $V$-band light curve with Equation (4) from \citet{2010ApJ...715..833O}, we obtain the middle of the transition phase as $t_{\rm{PT}}$ = 95 d. Using the $V-I$ colour at 65 d and correcting for the Galactic reddening, we obtain ${A_v}(host)$ = 0.05$\pm$0.21 mag. Thus we adopt the extinction $E(B-V)_{\rm tot}$ = 0.04 mag for \sn.

\begin{figure}
\centering
\includegraphics[width=0.5\textwidth]{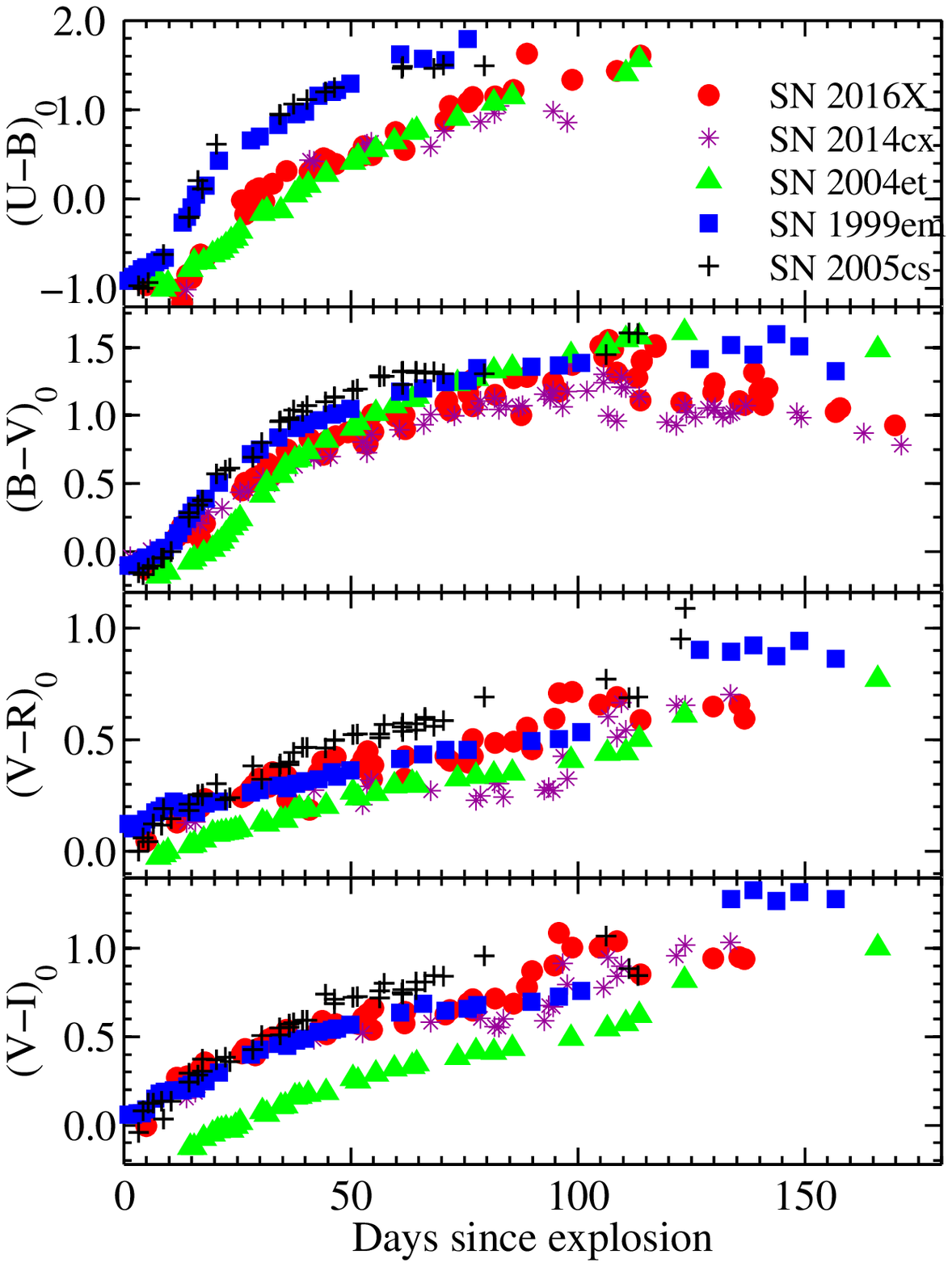}
\caption{The colour-curve evolution of \sn, along with that of other well-studied SNe IIP (SNe 1999em, 2004et, 2005cs, and 2014cx). All the colours have been corrected for both the Galactic and host-galaxy reddening.}
\label{fig:colour}
\end{figure}

In Figure \ref{fig:colour}, we show the reddening corrected $(U-B)_0, (B-V)_0, (V-R)_0$, and $(V-I)_0$ colour curves of \sn\, together with those of a few comparison SNe IIP. The colour evolution of \sn\ shows similar trend with that of other SNe IIP. At early time, the $(U-B)_0$ and $(B-V)_0$ colours are quite blue and they evolve towards redder colours rapidly as a result of faster expansion and cooling of the ejecta. In comparison, the $(V-R)_0$ and $(V-I)_0$ colours evolve more slowly with a rate of $<$ 0.5 mag in 30 days. During the plateau phase ($\sim$30--110) days, the $(U-B)_0$ and $(B-V)_0$ colours become progressively red by $\sim$ 1 mag as the cooling rate decreases, while $(V-R)_0$ and $(V-I)_0$ colours show little change. The $(B-V)_0$ colour shows a peak during the transitional phase around t $\sim$ +110 days, which is also visible in other SNe IIP. In the nebular phase ($>$ 120 days), the $B-V$ colour becomes gradually bluer, similar to that of SN 1999em and SN 2014cx.

\subsection{Bolometric Light Curve} \label{subsec:bolo}
Due to the lack of near-infrared observations, we calculated the quasi-bolometric luminosity of \sn\ following the same method as described in \citet{2015ApJ...807...59H}. After corrections for the line-of-sight extinction, the broadband magnitudes were converted into fluxes at the effective wavelength when $V$-band observations were available. The data in other bands, if not obtained, were estimated by interpolating the observations on adjacent nights. The spectral energy distribution (SED) were integrated, and the observed fluxes were converted to luminosity with the Tully-Fisher distance from \cite{2014MNRAS.444..527S}.

\begin{figure}
\centering
\includegraphics[width=0.5\textwidth]{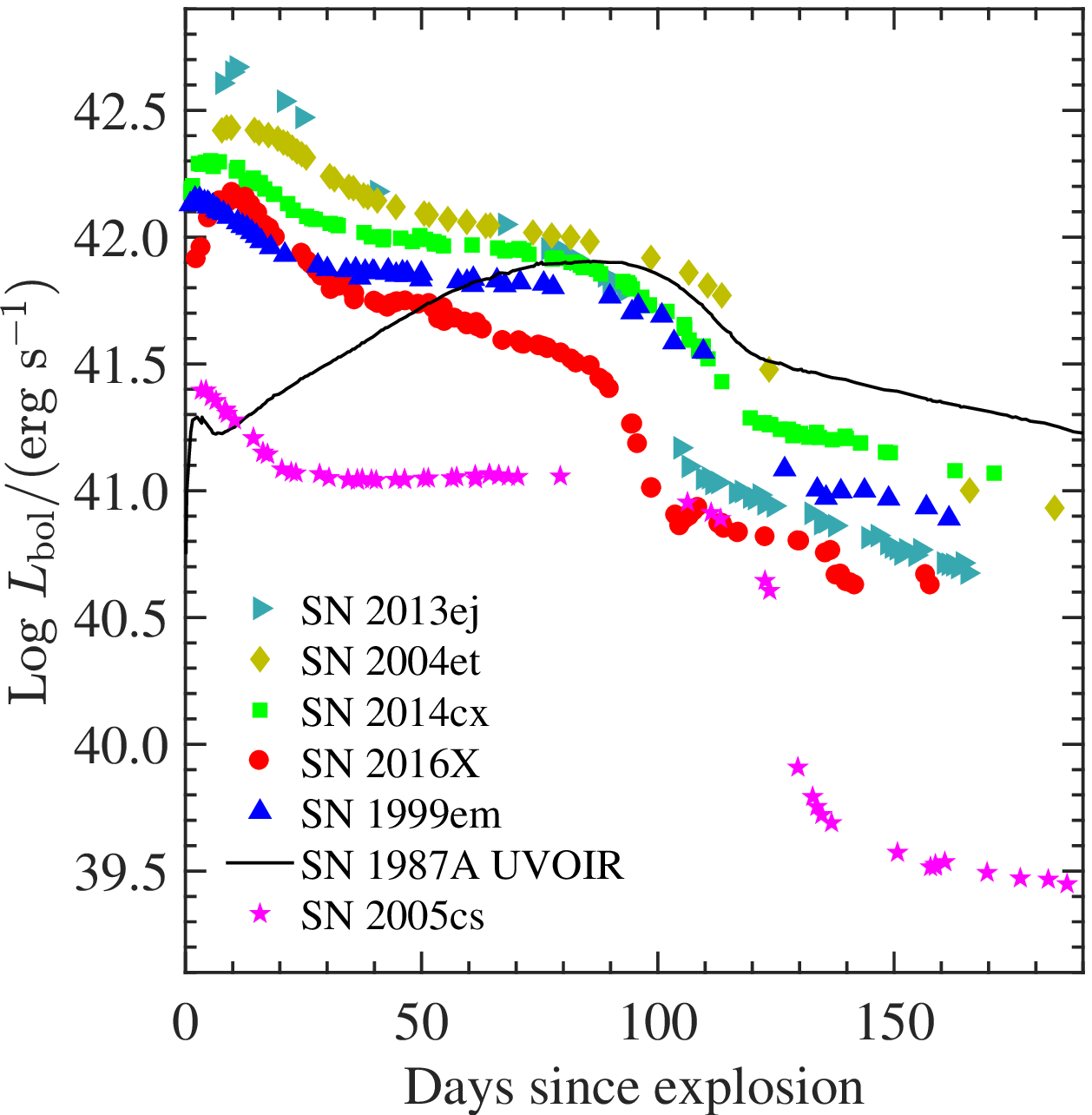}
\caption{The quasi-bolometric light curve of \sn\ compared with that of a few well-studied SNe IIP.}
\label{fig:bolo}
\end{figure}

Figure \ref{fig:bolo} shows the quasi-bolometric UV+optical ($UBVRI$) light curve of \sn, compared with that of some representative SNe IIP. The peak luminosity is estimated to be as log $L_{\rm bol}$ = 42.17 \ergs. Note that the calculations of the quasi-bolometric light curves still suffer large uncertainties in the distance modulus. The plateau luminosity of \sn\ is not constant but shows a monotonic decline up to t $\sim$ 90 days after explosion, which is similar with SN 2004et and SN 2013ej. The decline rate during the plateau phase is faster than other normal SNe IIP but comparable to the fast-declining type IIP SN 2013ej. The tail luminosity is lower than that of comparison SNe IIP except for the sub-luminous SN 2005cs, indicating that a relatively small amount of \nickel\  was synthesized in the explosion. Using the least-square fitting, the decline rate at the nebular phase is estimated to be 0.6 mag (100d)$^{-1}$.

\vspace{3mm}

\begin{figure}
\centering
\includegraphics[width=0.5\textwidth]{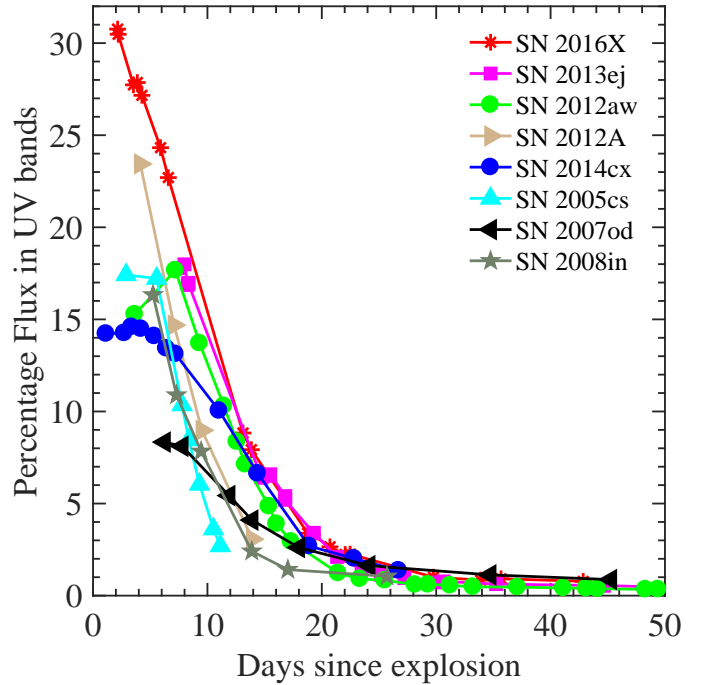}
\caption{Temporal evolution of the UV contribution of SN 2016X in the first two months, compared with some well-studied SNe IIP (including SNe 2005cs, 2007od, 2008in, 2012A, 2012aw, 2013ej, and 2014cx). The comparison data are extracted from Swift Optical/Ultraviolet Supernova Archive.}
\label{fig:uvratio}
\end{figure}

For the bolometric luminosity, the UV flux has a significant contribution in the early time ($\le$ 30 d), as shown in Figure \ref{fig:uvratio}. The UV contribution can reach $\gtrsim$30\% for \sn, which is much higher than other comparison SNe IIP (i.e., $\sim$15\%). Prominent UV emission is also in agreement with the higher temperature and larger progenitor radius derived for \sn\ in Section \ref{subsec:radi}. After about one month, the UV contribution becomes marginally important for most SNe IIP when entering into the plateau phase. Note that the above calculations of UV fraction may suffer from the uncertainties in dust extinctions applied for different SNe IIP. 

\vspace{1mm}

\section{Spectroscopic Analysis} \label{sec:spec}
\subsection{Evolution of Optical Spectra} \label{subsec:spe}
\begin{figure*}
\centering
\includegraphics[width=0.8 \textwidth]{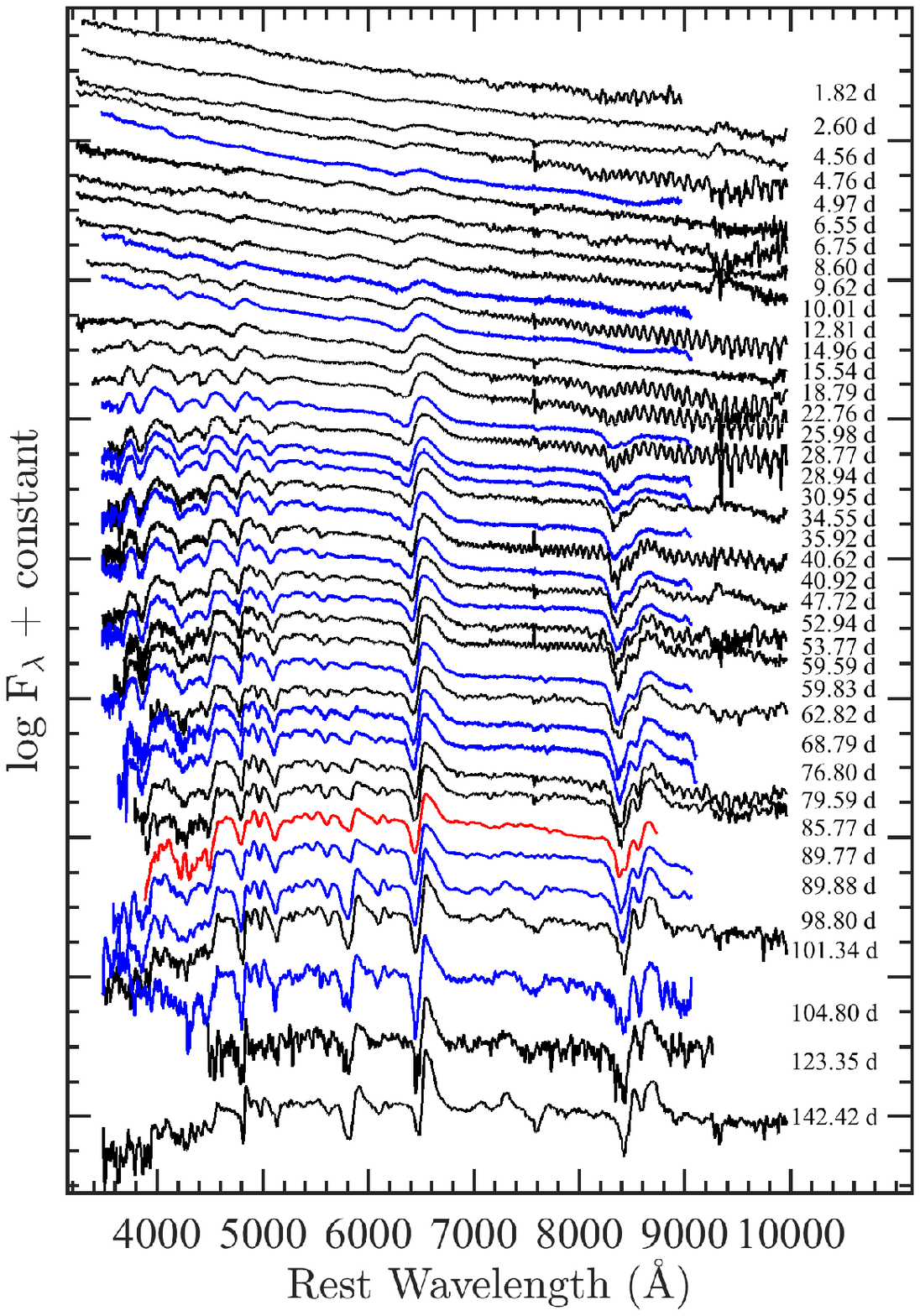}
\caption{The spectra sequence of \sn\ . The spectra obtained with the LCO, LJ 2.4-m telescope, and the Xinglong 2.16-m telescope are shown in black, blue and red curves, respectively. The phase relative to the explosion date (MJD = 57,405.92) is shown on the right of each spectrum.}  \label{fig:spec}
\end{figure*}

A total of 40 optical spectra of \sn\ covering the phase from +2 d to +140 d after the explosion are displayed in Figure \ref{fig:spec}. The phases marked in the plot are relative to the explosion date estimated in \S 3.1. All spectra have been corrected for the recession velocity of the host galaxy (1321$\pm$2 \kms)\footnote{http://leda.univ-lyon1.fr/}. The main spectral features are identified in previous studies for SNe IIP \citep{2002PASP..114...35L, 2004MNRAS.347...74P}, and are also marked in Fig \ref{fig:speccomp}.

The first spectrum, taken at less than 2 days after explosion, shows a featureless blue continuum, consistent with a very young event of core-collapse explosion. The blue continuum indicates that the photosphere has a temperature that is above 10$^4$ K. At t $\approx$ 2.6 d, shallow hydrogen Balmer lines, and He {\sc i} \ld 5876 lines with broad P-Cygni profiles become visible. The blue wing of \ha\ absorption indicates that the expansion velocity can reach up to $\sim$18,000 \kms. A double P-Cygni absorption of \ha\ appears in the t=+8d spectrum (see Fig \ref{fig:speccomp}(a)), and disappears after one month since explosion. The high-velocity feature is also reported in other SNe IIP, which might be due to Si {\sc ii} \ld 6355.

After two weeks since explosion (t $\ge$ 15 d), the He {\sc i} feature vanishes and is replaced by Na {\sc i} line at the similar position. Apart from hydrogen Balmer lines, O {\sc i} \ld 7774, Ca {\sc ii} H $\&$ K (\ld 3934, 3968), Ca {\sc ii} NIR triplet (\ld 8498, 8542, 8662), and Fe {\sc ii} multiplets are also clearly seen in the spectra. During the photospheric phase, the spectra turn progressively redder, and a number of narrow metal lines (Fe {\sc ii}, Ti {\sc ii}, Sc {\sc ii}, Ba {\sc ii}, Mg {\sc ii}, et al.) emerge in the spectra. These features grow progressively stronger and dominate the spectra over time.

After $\sim$90 days, the continuum flattens, and the spectra become dominated by emission lines, meaning that the SN enters into the nebular phase. The \ha\ emission profile shows a weak asymmetric feature (also seen in Figure \ref{fig:speccomp}c). The asymmetric feature has been commonly observed in a few SNe IIP (\ie, SNe 1999em, 2004dj, and 2013ej), and might result from interaction with circumstellar medium, asymmetry in the line-emitting region \citep{2002PASP..114...35L}, or bipolar \nickel\ distribution in a spherical envelope \citep{2006AstL...32..739C}. The subsequent spectra show permitted lines due to metals, when the outer ejecta became optically thin. And the spectra are characterized by the presence of forbidden lines [O {\sc i}] \ld\ld 6300, 6364 and [Ca {\sc ii}] \ld\ld 7291, 7324.

\subsection{Comparison with Other SNe IIP}
\begin{figure}
\centering
\includegraphics[width=0.5\textwidth]{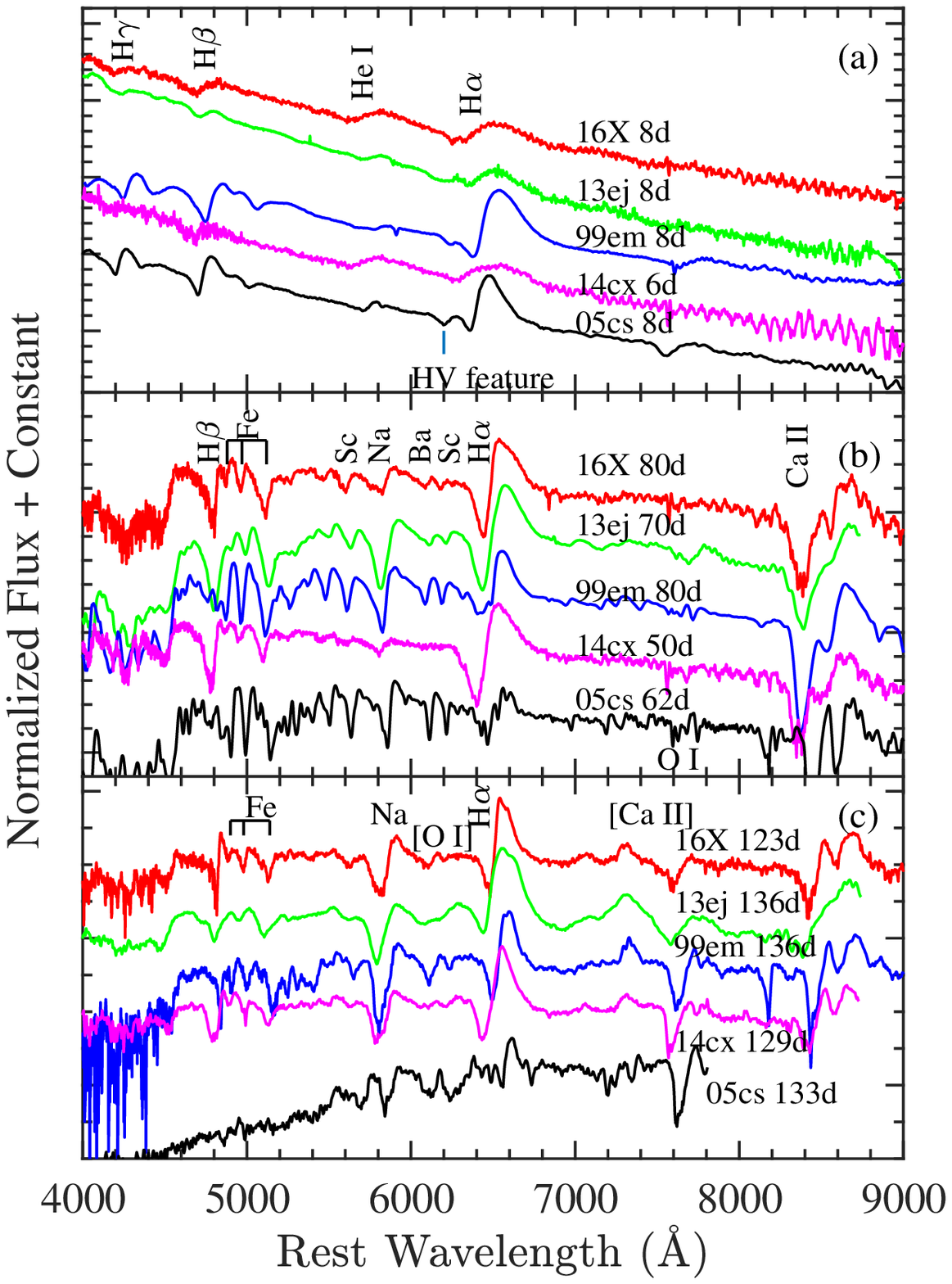}
\caption{Spectral comparison  of \sn\ with those of other well-studied SNe IIP at early (1 week), plateau (80 d), and nebular (130 d) phases.} \label{fig:speccomp}
\end{figure}

In Figure \ref{fig:speccomp}, we compare the spectra evolution of \sn\ to a few other SNe IIP at similar phases, \ie\ the early phase at one week, the plateau phase at 2 months, and the nebular phase at 4 months after explosion. \sn\ shows similarities with these comparison SNe IIP (especially SN 2013ej and SN 2014cx) in the spectral evolution. In the early phase, the spectrum of \sn\ shows weaker and broader profiles of Balmer lines and He {\sc I} line compared to SN 1999em and SN 2005cs. During the plateau phase, the spectra of \sn\ and the comparison objects are dominated by metal lines, including Fe {\sc ii}, Ti {\sc ii}, Sc {\sc ii}, Ba {\sc ii}, and Mg {\sc ii} etc. The forbidden lines such as [O I] and [Ca II] emerge in the spectra when the SNe enter into the nebular phase. In comparison, SN 2005cs shows much narrower absorption features and redder continuum at this phase. 

\subsection{Expansion Velocities} \label{subsec:vel}
\begin{figure}
\centering
\includegraphics*[width=8.5cm]{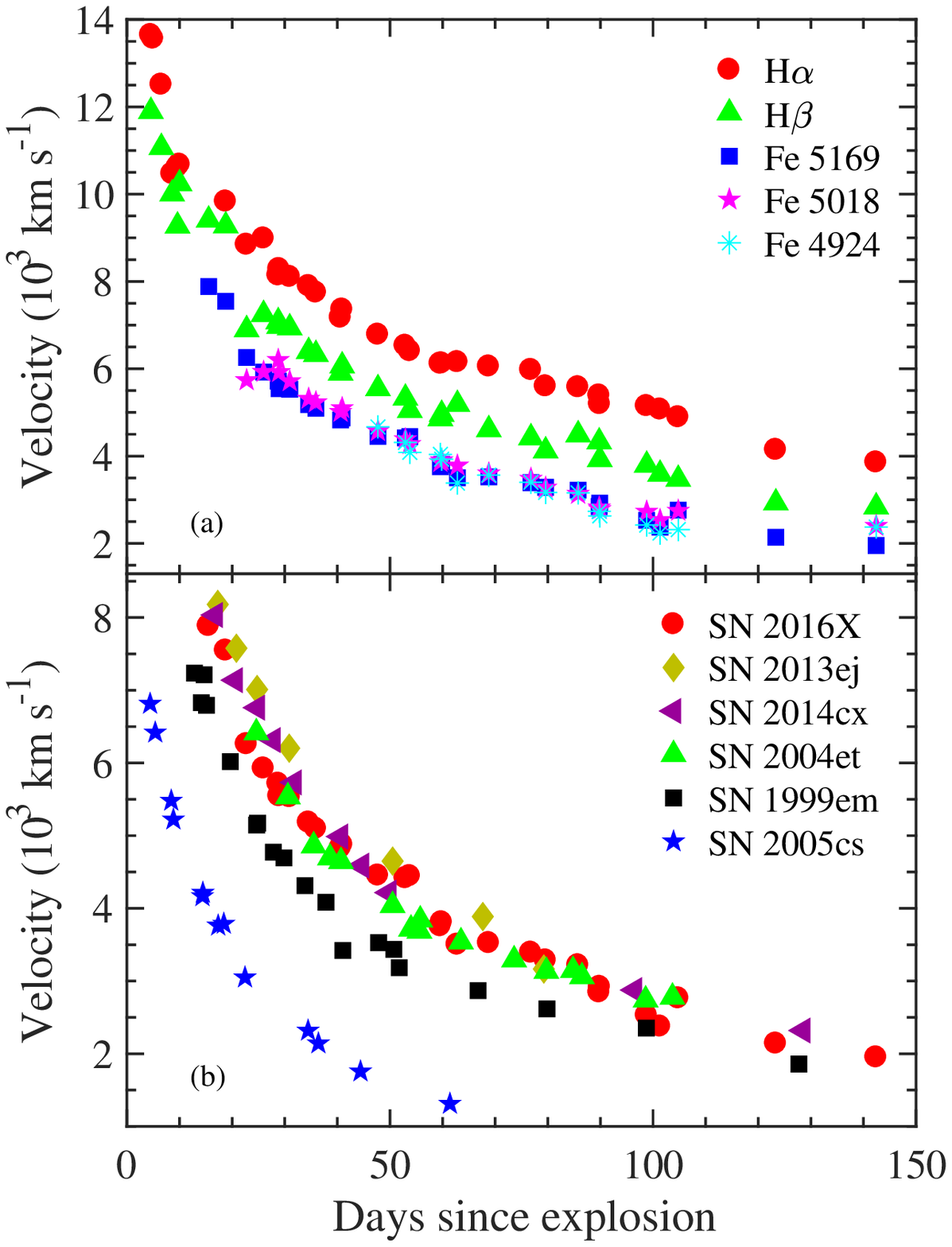}
\caption{Upper: The velocity evolution of \ha\, \hb\ and Fe {\sc ii} lines. Lower: Comparison of evolution of photospheric velocity (measured by Fe \ld 5169) of \sn\ with some well-studied type II-P SNe such as SN 1999em \citep{2002PASP..114...35L}, SN 2004et \citep{2006MNRAS.372.1315S}, SN 2005cs \citep{2009MNRAS.394.2266P}, SN 2013ej \citep{2015ApJ...807...59H}, and SN 2014cx \citep{2016ApJ...832..139H}.}  \label{fig:velo}
\end{figure}

The measurement of the ejecta velocities and comparison with that of other SNe IIP are presented in this subsection. The expansion velocities of hydrogen and metal lines are measured by using SPLOT in IRAF to locate the absorption minima. The upper panel of Figure \ref{fig:velo} shows the line velocities of \ha, \hb, Fe II \ld 5169, 5018 and 4924. During the first week after the explosion, the expansion velocity of hydrogen is above 10,000 \kms and it declines very rapidly. Later on, the velocity then declines in an exponential trend over time. The velocity of Fe II lines, which is a good indicator of photospheric velocity, is always lower than that of hydrogen lines and it decreases below 3,000 \kms after 90 d. This can be explained by that the Fe II lines are formed in the inner-layers with larger optical depths.

In the lower panel of Figure \ref{fig:velo}, we compare the velocity evolution of Fe \ld 5169 between \sn\ and other SNe IIP. It is obvious that the velocity of \sn\ higher than SN 1999em (by $\sim$ 1,000 \kms) and SN 2005cs (by $\sim$ 3,000 \kms), and close to SN 2013ej, SN 2004et, and SN 2014cx.

\section{Discussion} \label{sec:discuss}

\subsection{Nickel Mass} \label{subsec:nick}
The amount of \nickel\ synthesized in the explosion of SNe IIP can be estimated by the luminosity of their late-time light curves. In the nebular phase, the light curve is powered by the radioactive decay of \nickel\ to \cobalt\ and \cobalt\ to \iron, with $e$-folding time of 8.8 d and 111.26 d, respectively. The mass of \nickel\ of SN 1987A has been accurately determined to be 0.075 $\pm$ 0.005 \msun\ \citep{1996snih.book.....A}. We adopt a linear least square fit to the nebular luminosity during the phase 120--160 d, and obtain $L$(\sn)/ $L$(SN 1987A) = 0.43 at 140 d, from which we derive $M(^{56}{\rm Ni}) = 0.032\pm$0.006 \msun. 

Assuming that the $\gamma$ photons produced from the \nickel\ to \iron\ are fully thermalized, the \nickel\ mass can be also estimated from the tail luminosity. Using Equation (2) in \citet{2003ApJ...582..905H}, we estimate the \nickel\ mass to be $M( ^{56}{\rm Ni}) = 0.034 \pm$ 0.005 \msun.

\cite{2003A&A...404.1077E} found a tight correlation between the \nickel\ mass and a steepness parameter of the $V$-band light curve at the transitional phase. For \sn\, we fit the $V$-band light curve and estimate the steepness parameter $S$ as 0.099 mag day$^{-1}$ at the epoch of inflection $t_i = 92$ day. Using the empirical relation ($\log\,M( ^{56}{\rm Ni}) = -6.2295\,S -0.8147$), the mass of \nickel\ for \sn\ is estimated to be 0.037$\pm$0.003 \msun. This value is consistent with that derived from the tail luminosity. Therefore, the average value of \nickel\ mass is taken as 0.034 $\pm$ 0.006 \msun.

\subsection{Properties of Progenitor} \label{subsec:radi}
For CC~SNe, shortly after the shock breakout, the shock-heated stellar envelope cools down due to the outward expansion. The timescale of cooling depends mainly on the initial radius of the progenitor, opacity, and gas composition. And the early light curves of SNe are dominated by the radiation from the expanding envelope. Some simple analytic expressions have been developed to describe the properties of the emitted radiation and are used to constrain the progenitor radius (\eg\ \citealt{2011ApJ...728...63R, 2011ApJ...729L...6C, 2017ApJ...838..130S}). For example, progenitors with larger radius (i.e., RSG with 500--1000 \rsun) stay at higher temperature and cool down at a slower pace than those with smaller radius (i.e., BSG with 50--100 \rsun), as indicated by the expression \mbox{$ T_{\text{ph}}(t) =1.6 \, f_{\rho}^{-0.037}
    \frac{E_{51}^{0.027} R_{*,13}^{1/4} }{(M/M_{\odot})^{0.054} \kappa^{0.28}_{0.34}}
    t_5^{-0.45} $ eV,} where $f_{\rho}$ represents density profile, $E_{51}$ is the energy in units of 10$^{51}$ erg, $R_{*,13}$ is the radius in units of $10^{13}$ cm, $\kappa_{0.34}$ is the opacity in units of 0.34 cm$^2$ g$^{-1}$, and $t_5$ is time in units of 10$^5$ s.
    
\begin{figure}
\centering
\includegraphics*[width=8.5cm]{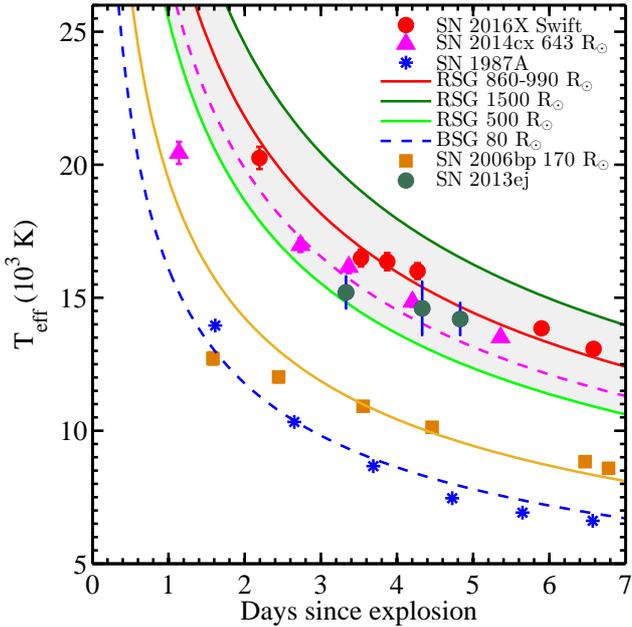}
\caption{Radius estimates using the prescription from \citet{2011ApJ...728...63R}. The best fit for \sn\ is 860 -- 990 \rsun. Over-plotted are the comparison objects SN 1987A, 2006bp, 2013ej and 2014cx \citep{2014MNRAS.438L.101V, 2016ApJ...832..139H}.} \label{fig:radius}
\end{figure}

Thanks to the timely follow-up observations from the \emph{Swift} UVOT, we are able to better construct the spectral energy distribution and estimate the corresponding blackbody temperature (cooling phase of the shock breakout) for \sn\ in the early phase. This allows us to constrain its progenitor radius by fitting to the temperature evolution. Adopting an optical opacity of 0.34 ${cm}^2 g^{-1}$ and a RSG density profile $f_{\rho}=0.13$ in the Eq.(13) from \citet{2011ApJ...728...63R}, we yield an initial radius of 860--990 \rsun\ for the progenitor of \sn~, as shown in Figure \ref{fig:radius}. We also overplot the temperature evolution and the best-fit progenitor radius for SN 1987A, SN 2006bp, SN 2013ej, and SN 2014cx. One can see that \sn\ has an apparently large progenitor in comparison with other SNe IIP.  

Using the SuperNova Explosion Code (SNEC, \citealt{2015ApJ...814...63M}), \citet{2016ApJ...829..109M} find that the early properties of the light curves of SNe IIP depend sensitively on the radius of the progenitor star, with a relationship between the $g$-band rise time and the radius at the time of explosion (i.e., log $R [R_{\odot}]$ = 1.225 log $t_{\rm rise}$ [day] + 1.692). We also use this relation to estimate the size of the progenitor star. For \sn\ , the $g$-band rise time is estimated as 10.60$\pm$0.40 days, which leads to an estimate of 890$\pm$40 \rsun\ for the progenitor of \sn\ . This analysis, together with the result from shock breakout cooling, favours that \sn\ has a larger progenitor with a radius up to $\sim$ 900-1000 \rsun.

\begin{figure}
\centering
\includegraphics*[width=8.5cm]{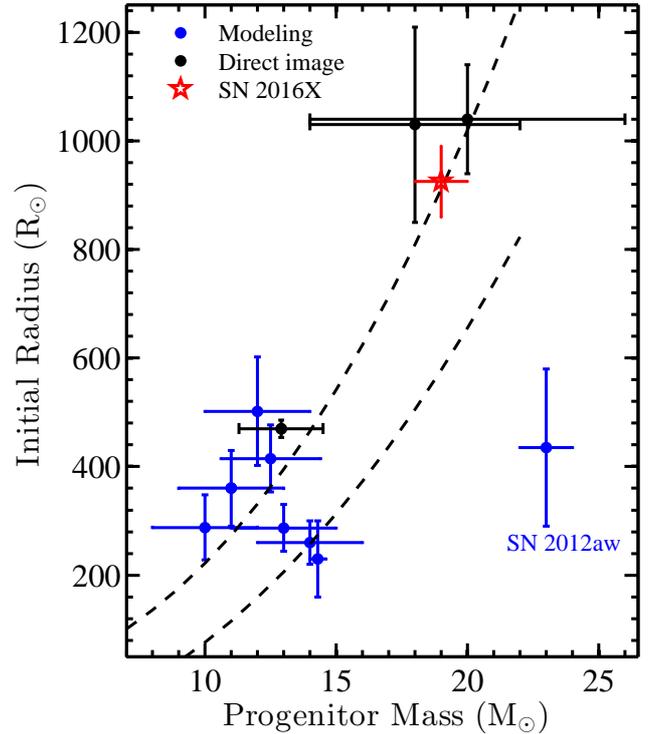}
\caption{The progenitor mass and radius for a sample of SNe IIP. The blue dots represent the estimates using hydrodynamic modeling, while the black dots are results from analysis of the pre-explosion images (see the reference from Table \ref{tab:massnrad}). Dashed lines represent the mass--radius relation derived from the Galactic (upper) and Magellanic-cloud (lower) RSGs, respectively \citep{2015MNRAS.451.2212G}.} \label{fig:mrad}
\end{figure}

Based on the RSG sample in the Milky Way and Magellanic Clouds (MC) \citep{2005ApJ...628..973L, 2006ApJ...645.1102L}, \citet{2015MNRAS.451.2212G} found that there is a general tendency that the more massive RSG stars have larger radius sizes. A tight mass--radius relation can be obtained for the RSG stars in the Milky Way, \ie\ $R/R_{\odot}=1.4(M/M_{\odot})^{2.2}$, as shown in Figure \ref{fig:mrad} (see the dashed line). This relation gives a rough estimate of 18.5--19.7 \msun\ for the progenitor of  \sn. In this plot, we also show the progenitor mass and radius estimated from photospheric cooling/hydrodynamic analysis and HST archive images for a sample of SNe IIP. We notice that mass and radius of these SNe IIP seem to follow that of the Galactic or MC RSGs, except that the hydrodynamic method gives a larger mass and a smaller radius for SN 2012aw. Table \ref{tab:massnrad} listed the details of these estimates and the references. For comparison, we overplot the mass--radius relation derived from red supergiants in the MCs (see the dotted line). Given a radius, the star will have a larger mass for lower metallicity. This can be explained with that more metal-poor stars usually lose their mass at a lower efficiency. 

As the host galaxy of \sn\ UGC 08041 is a late-type Sd galaxy and the sn locates at its outskirts, it is possible that the progenitor of \sn\ has a relatively lower metallicity. Considering this effect and hence the possible mass loss of the progenitor star before the explosion, the mass range we derived for the progenitor of \sn\ should be a lower limit. Along with SN 2012aw and SN 2012ec, the high mass derived for the progenitor of \sn\ indicates that RSGs with an initial mass around or above 20.0 \msun\ could lead to an explosion of type IIP SN. 

\section{Summary} \label{sec:summ}
In this paper, we present the ultraviolet/optical photometry and low resolution spectroscopic observations for the type IIP \sn\ up to 180 days after explosion. The high-quality UV/optical data allow us to place interesting constraints on the observational properties of \sn\ and its progenitor. A brief summary of our results are listed below. 

The Swift UVOT data reveals the presence of prominent UV emissions at just only 2 days before the primary UV peaks, which is very likely related to the shock breakout of very massive stars. For \sn\, the UV contribution to the total flux can reach $\gtrsim$30\% for \sn\ in the early phase, while the typical value is $\sim$15\%. In particular, this supernova is found to have a very long rise time before reaching the maximum light, i.e., 12.6$\pm$0.5 days in the $R$ band, in contrast to $\sim$7.0 days for normal SNe IIP.
The photometric and spectral evolution is overall similar to SN 2013ej. 

Using the early-time temperature evolution inferred from the \emph{Swift} UV photometry, we derived an initial radius of 860--990 \rsun\ for the progenitor of \sn. The long $g$-band rise time of \sn\ also indicates a large progenitor radius of $\sim$ 890 \rsun\ according to the rise time -- radius relation from the SNEC. Based on the mass -- radius relation of the Galactic RSG, we also obtain a rough mass estimate of 18.5--19.7 \msun\ for the progenitor of \sn, which provides further evidence that massive stars with an initial mass up to 19-20 \msun could also produce an explosion of type IIP supernova.

\section*{Acknowledgements}
We acknowledge the support of the staff of the Lijiang 2.4m and Xinglong 2.16m telescope. Funding for the LJT has been provided by Chinese Academy of Sciences and the People's Government of Yunnan Province. The LJT is jointly operated and administrated by Yunnan Observatories and Centre for Astronomical Mega-Science, CAS. This work is supported by the National Natural Science Foundation of China (NSFC grants 11178003, 11325313, and 11633002). This work was also partially supported by the Collaborating Research Program (OP201702) of the Key Laboratory of the Structure and Evolution of Celestial Objects, Chinese Academy of Sciences. This work makes use of observations from Las Cumbres Observatory.  DAH, CM, and GH are supported by the US National Science Foundation grant 1313484. Support for IA was provided by NASA through the Einstein Fellowship Program, grant PF6-170148. The work made use of Swift/UVOT data reduced by P. J. Brown and released in the Swift Optical/Ultraviolet Supernova Archive (SOUSA). SOUSA is supported by NASA's Astrophysics Data Analysis Program through grant NNX13AF35G. J.-J. Zhang is supported by the NSFC (grants 11403096, 11773067), the Key Research Program of the CAS (Grant NO. KJZD-EW- M06), the Youth Innovation Promotion Association of the CAS, and the CAS ``Light of West China'' Program. T.-M. Zhang is supported by the NSFC (grants 11203034).





\begin{table*}
\caption{Photometric Standard Stars in the Field of \sn\ ($1\sigma$ Uncertainties). } \label{tab:standstar}
\begin{tabular}{cccccccc}
\hline \hline
 Star& $\alpha_{\rm J2000}$&       $\delta_{\rm J2000}$& $U$ & $B$&  $V$&  $R$&  $I$\\
   ID&          (h m s)      &(\arcdeg\ \arcmin\ \arcsec)  &(mag)&(mag)&(mag)&(mag)&(mag)\\
  \hline
  1    &  12:55:10.458 &  0:09:19.79 & 18.58(03) &18.38(04) &17.58(03) &17.11(03) &16.66(06)   \\
  2    &  12:55:22.596 &  0:08:42.00 & 14.47(02) &13.91(03) &13.47(02) &13.10(05) &12.39(09)   \\
  3    &  12:55:26.850 &  0:04:14.02 & 18.81(04) &17.78(06) &16.62(03) &15.95(04) &15.33(09)   \\
  4    &  12:55:21.901 &  0:04:33.88 & 17.92(03) &17.59(05) &16.77(03) &16.29(04) &15.81(07)   \\
  5    &  12:55:11.357 &  0:04:50.16 & 18.38(03) &17.24(07) &15.72(04) &14.78(08) &13.72(16)  \\
  6    &  12:55:00.181 &  0:05:08.91 & 16.93(02) &17.05(04) &16.48(03) &16.14(03) &15.78(05)   \\
  7    &  12:55:16.953 &  0:04:05.97 & 20.37(08) &19.44(07) &17.94(04) &16.92(10) &15.56(21)  \\
  8    &  12:55:26.305 &  0:03:09.69 & 16.20(02) &15.98(04) &15.30(03) &14.90(03) &14.49(06)   \\
  9    &  12:55:14.299 &  0:03:17.25 & 19.16(04) &18.97(04) &18.27(03) &17.85(03) &17.41(06)   \\
  10   &  12:55:17.333 &  0:03:16.92 & 15.72(02) &15.68(04) &15.01(03) &14.60(03) &14.15(06)   \\
 \hline
\end{tabular}
\end{table*}

\begin{table*}
\centering
\caption{Optical photometry from TNT.} \label{tab:phot_tnt}
\begin{tabular}{c c c c c c c c}
\hline \hline
 UT Date &    MJD       & Phase$^{a}$  &  $U$    &     $B$     & $V$       & $R$    &    $I$ \\
(yy/mm/dd)&             & (day)       & (mag)   &    (mag)    &    (mag)   &   (mag)&   (mag)   \\
\hline
2016 Jan. 30 & 57417.705 &   11.79 & 13.28(01) &14.26(03) & 14.11(01) & 13.95(01) & 13.78(04)    \\
2016 Jan. 31 & 57418.755 &   12.84 & 13.19(01) &14.32(02) & 14.09(02) & 13.89(03) & 13.77(02)    \\
2016 Feb. 01 & 57419.885 &   13.97 & 13.46(01) &14.28(04) & 14.10(04) & 13.89(05) & 13.75(05)    \\
2016 Feb. 02 & 57420.710 &   14.79 & 13.61(02) &14.44(03) & 14.16(04) & 14.06(05) & 93.94(04)    \\
2016 Feb. 04 & 57422.775 &   16.86 & \nodata   &14.44(03) & 14.21(03) & 13.99(03) & 13.82(02)    \\
2016 Feb. 05 & 57423.870 &   17.95 & 13.88(02) &14.51(03) & 14.26(04) & 14.00(02) & 13.84(04)    \\
2016 Feb. 14 & 57432.680 &   26.76 & 14.82(01) &14.96(02) & 14.41(01) & 14.13(03) & 13.91(02)    \\
2016 Feb. 15 & 57433.885 &   27.97 & 14.97(02) &15.01(03) & 14.46(01) & 14.15(04) & 13.99(02)    \\
2016 Feb. 16 & 57434.690 &   28.77 & 15.03(02) &15.07(02) & 14.49(03) & 14.16(03) & 13.99(04)    \\
2016 Feb. 17 & 57435.695 &   29.78 & 15.24(02) &15.09(03) & 14.49(04) & 14.13(05) & 13.99(04)    \\
2016 Feb. 19 & 57437.885 &   31.97 & \nodata   &15.19(03) & 14.53(03) & 14.21(02) & 13.98(03)   \\
2016 Feb. 20 & 57438.675 &   32.76 & 15.40(03) &15.20(04) & 14.60(04) & 14.21(04) & 14.04(04)    \\
2016 Feb. 22 & 57440.705 &   34.79 & \nodata   &15.24(01) & 14.58(02) & 14.20(01) & 14.03(01)    \\
2016 Feb. 23 & 57441.885 &   35.97 & \nodata   &15.33(03) & 14.54(03) & 14.28(04) &   \nodata    \\
2016 Mar. 01 & 57448.845 &   42.93 & 15.93(04) &15.65(03) & 14.74(04) & 14.36(06) & 14.14(05)    \\
2016 Mar. 02 & 57449.690 &   43.77 & 15.96(05) &15.51(03) & 14.72(04) & 14.29(04) & 14.07(05)    \\
2016 Mar. 03 & 57450.670 &   44.75 & \nodata   &15.52(02) & 14.67(03) & 14.28(03) & 14.09(03)   \\
2016 Mar. 05 & 57452.625 &   46.71 & 15.98(05) &15.56(05) & 14.67(05) & 14.22(04) & 14.03(05)    \\
2016 Mar. 10 & 57457.635 &   51.72 & 16.12(03) &15.61(05) & 14.66(05) & 14.26(05) & 14.03(04)    \\
2016 Mar. 11 & 57458.645 &   52.73 & 16.12(05) &15.60(04) & 14.75(02) & 14.31(03) & 14.08(03)    \\
2016 Mar. 12 & 57459.630 &   53.71 & 16.26(04) &15.70(04) & 14.87(04) & 14.40(04) & 14.18(04)    \\
2016 Mar. 13 & 57460.625 &   54.71 & 16.23(05) &15.70(04) & 14.66(07) & 14.30(06) & 14.05(04)    \\
2016 Mar. 20 & 57467.715 &   61.80 & 16.43(07) &15.85(03) & 14.81(04) & 14.46(05) & 14.17(05)    \\
2016 Mar. 29 & 57476.685 &   70.77 & 17.02(05) &16.13(03) & 15.00(04) & 14.54(03) & 14.30(04)    \\
2016 Mar. 30 & 57477.695 &   71.78 & 17.18(07) &16.11(04) & 15.04(03) & 14.60(06) & 14.32(04)    \\
2016 Apr. 03 & 57481.750 &   75.83 & 17.35(05) &16.24(02) & 15.04(01) & 14.61(04) & 14.28(03)    \\
2016 Apr. 04 & 57482.750 &   76.83 & 17.39(09) &16.22(03) & 15.11(03) & 14.59(03) & 14.33(05)    \\
2016 Apr. 13 & 57491.750 &   85.83 & 17.80(07) &16.55(03) & 15.23(03) & 14.71(03) & 14.48(03)    \\
2016 Apr. 16 & 57494.750 &   88.83 & 18.40(14) &16.75(03) & 15.42(04) & 14.84(04) & 14.58(02)    \\
2016 Apr. 22 & 57500.750 &   94.83 & \nodata   &\nodata   & 15.87(05) & 15.25(05) & 14.91(03)   \\
2016 Apr. 23 & 57501.750 &   95.83 & \nodata   &17.41(10) & 16.18(04) & 15.44(03) & 15.03(03)    \\
2016 May 06 & 57514.500  &  108.58 & 19.59(34) &18.14(05) & 16.77(03) & 16.05(04) & 15.67(04)    \\
2016 May 07 & 57515.750  &  109.83 & \nodata   & \nodata  &  \nodata  & 16.14(17) & 15.76(17)    \\
2016 Jun. 02 & 57541.500 &  135.58 & \nodata   &18.36(05) & 17.23(04) & 16.53(03) & 16.20(03)    \\
\hline
\end{tabular}
\begin{flushleft}
 $^{a}$ Relative to the explosion date, MJD = 57,405.92.\\
\end{flushleft}
\end{table*}

\begin{table*}
\centering
\caption{Optical photometry from Lijiang 2.4-m Telescope.} \label{tab:phot_LJ}
\begin{tabular}{c c c c c c c c}
\hline \hline
 UT Date &    MJD       & Phase$^{a}$  &  $U$    &     $B$     & $V$       & $R$    &    $I$ \\
(yy/mm/dd)&             & (day)       & (mag)   &    (mag)    &    (mag)   &   (mag)&   (mag)   \\
\hline
2016 Jan. 23 & 57410.91 &    4.99 & 13.36(02) & 14.29(04) & 14.38(03) & 14.31(04) & 14.33(02)  \\
2016 Jan. 28 & 57415.92 &   10.00 & 13.23(02) & 14.02(05) & 14.04(04) & 14.09(04) & 13.81(05)  \\
2016 Feb. 02 & 57420.88 &   14.96 & 13.53(06) & 14.38(02) & 14.16(02) & 13.93(03) & 13.83(03)  \\
2016 Feb. 04 & 57422.83 &   16.91 & 13.90(17) & 14.49(09) & 14.36(08) & 14.16(09) & 13.95(05)  \\
2016 Feb. 13 & 57431.92 &   26.00 & 14.88(05) & 14.86(05) & 14.37(02) & 14.10(01) & 13.90(08)  \\
2016 Feb. 16 & 57434.85 &   28.93 & 15.19(02) & 15.06(02) & 14.48(02) & 14.17(02) & 14.02(01)  \\
2016 Feb. 18 & 57436.90 &   30.98 & 15.27(03) & 15.27(03) & 14.69(03) & 14.23(03) & 14.06(01)  \\
2016 Feb. 23 & 57441.84 &   35.92 & 15.79(03) & 15.45(02) & 14.67(05) & 14.30(07) & 14.06(11)  \\
2016 Feb. 28 & 57446.83 &   40.91 & 15.83(07) & 15.49(08) & 14.62(05) & 14.41(09) &  \nodata   \\
2016 Mar. 02 & 57449.93 &   44.01 & 15.96(02) & 15.47(03) & 14.72(02) & 14.32(03) & 14.08(02)  \\
2016 Mar. 03 & 57450.84 &   44.92 & 15.96(04) & 15.49(02) & 14.69(03) & 14.26(03) & 14.07(03)  \\
2016 Mar. 11 & 57458.85 &   52.93 & 16.21(03) & 15.59(04) & 14.75(02) & 14.30(06) & 14.09(02)  \\
2016 Mar. 16 & 57460.91 &   54.99 & 16.36(13) & 15.78(06) & 14.86(04) & 14.45(03) & 14.14(03)  \\
2016 Mar. 18 & 57465.74 &   59.82 & 16.66(12) & 15.89(04) & 14.85(04) &  \nodata  &  \nodata   \\
2016 Mar. 20 & 57467.88 &   61.96 &  \nodata  & 15.86(01) & 14.92(02) & 14.46(02) & 14.21(02)  \\
2016 Apr. 04 & 57482.75 &   76.83 &  \nodata  & 16.39(03) & 15.09(04) & 14.63(04) & 14.37(02)  \\
2016 Apr. 09 & 57487.73 &   81.81 & 17.56(08) & 16.39(04) & 15.19(03) & 14.68(04) & 14.41(01)  \\
2016 Apr. 17 & 57495.79 &   89.87 &  \nodata  & 17.05(15) & 15.45(12) & 14.96(03) & 14.51(05)  \\
2016 Apr. 26 & 57504.72 &   98.80 & 19.36(16) & 18.00(06) & 16.58(04) & 15.83(04) & 15.51(06)  \\
2016 May  02 & 57510.68 &  104.76 &  \nodata  &\nodata    & 16.92(03) & 16.23(05) & 15.85(04)  \\
2016 May  11 & 57519.69 &  113.77 & 19.66(33) & 18.03(13) & 16.88(06) & 16.26(07) & 15.96(06)  \\
2016 May  27 & 57535.70 &  129.78 &  \nodata  & 18.30(06) & 17.08(03) & 16.40(03) & 16.07(02)  \\
2016 Jun. 03 & 57542.67 &  136.75 &  \nodata  & 18.29(14) & 17.17(09) & 16.55(05) & 16.17(05)  \\
\hline
\end{tabular}
\begin{flushleft}
  $^{a}$ Relative to the explosion date, MJD = 57,405.92.\\
   \end{flushleft}
\end{table*}

\begin{table*}
\centering
\caption{Optical photometry from LCO ($1\sigma$ Uncertainties).} \label{tab:phot_LCO}
\begin{tabular}{ccccccccc}
\hline \hline
UT Date &    MJD       & Phase$^a$     & $U$   &    $B$    & $V$       & $g$    &    $r$       & $i$    \\
(yy/mm/dd)&             & (day)       & (mag)   & (mag)  &    (mag)    &    (mag)   &   (mag)     & (mag)   \\
\hline
2016 Jan. 21 & 57408.340 &   2.421 & 13.723(023) & 14.711(042) & 14.818(035)  & 14.714(032) & 14.903(030) & 15.021(020)   \\
2016 Jan. 22 & 57409.360 &   3.441 & 13.718(052) & 14.504(042) & 14.675(041)  & 14.473(022) & 14.775(036) & 14.845(042)   \\
2016 Jan. 25 & 57412.275 &   6.356 & 13.615(012) & 14.199(044) & 14.186(052)  & 14.265(023) & 14.364(047) & 14.328(028)   \\
2016 Jan. 26 & 57413.315 &   7.396 & 13.276(023) & 14.128(030) & 14.083(032)  & 14.073(033) & 14.370(041) & 14.258(033)   \\
2016 Jan. 26 & 57413.340 &   7.421 & 13.259(027) & 14.130(027) & 14.070(026)  & 14.055(030) & 14.147(022) & 14.216(026)   \\
2016 Jan. 28 & 57415.085 &   9.166 & 13.300(044) & 14.088(038) & 14.162(035)  & 14.026(036) & 14.159(040) & 14.274(039)   \\
2016 Jan. 28 & 57415.235 &   9.316 & 13.276(041) & 14.162(032) & 14.198(034)  & 14.113(041) & 14.095(050) & 14.106(048)   \\
2016 Feb. 01 & 57419.245 &  13.326 & 13.509(052) & 14.288(041) & 14.109(029)  & 14.150(026) & 14.043(024) & 14.176(028)   \\
2016 Feb. 03 & 57421.090 &  15.171 & 13.565(016) & 14.348(041) & 13.991(026)  & 14.168(044) & 14.023(032) & 14.161(032)   \\
2016 Feb. 07 & 57425.095 &  19.176 & 14.284(018) & 14.515(058) & 14.261(050)  & 14.308(029) & 14.189(052) & 14.211(039)   \\
2016 Feb. 12 & 57430.655 &  24.736 & 14.460(035) & 14.790(056) & 14.407(054)  & 14.513(030) & 14.222(038) & 14.315(035)   \\
2016 Feb. 15 & 57433.995 &  28.076 & 14.995(073) & 15.067(042) & \nodata      & 14.631(043) & 14.269(040) & 14.387(043)   \\
2016 Feb. 19 & 57437.720 &  31.801 &   \nodata   & 15.178(053) & 14.490(038)  & 14.870(036) & 14.342(034) & 14.381(025)   \\
2016 Feb. 23 & 57441.565 &  35.646 & 15.356(037) & 15.323(044) & 14.682(039)  & 14.888(045) & 14.486(042) & 14.467(029)   \\
2016 Feb. 28 & 57446.035 &  40.116 & 15.689(032) & 15.426(032) & 14.657(039)  & 14.973(037) & 14.533(039) & 14.500(038)   \\
2016 Mar. 02 & 57449.765 &  43.846 & \nodata     &  \nodata    &  \nodata     & 14.985(036) & 14.548(053) & 14.516(025)   \\
2016 Mar. 08 & 57455.340 &  49.421 &   \nodata   & 15.628(062) & 14.710(042)  & 15.285(042) & 14.454(035) & 14.502(034)   \\
2016 Mar. 11 & 57458.665 &  52.746 & 15.881(057) & 15.621(045) & 14.787(029)  & 15.314(052) & 14.613(037) & 14.487(033)   \\
2016 Mar. 16 & 57463.075 &  57.156 & 16.141(063) & 15.769(052) & 14.799(017)  & \nodata     &  \nodata    &   \nodata     \\
2016 Mar. 18 & 57465.255 &  59.336 & 16.195(049) & 15.864(031) & 14.843(039)  & 15.345(028) & 14.572(030) & 14.532(022)   \\
2016 Mar. 21 & 57468.915 &  62.996 & 16.425(100) & 15.967(048) & 14.914(031)  & 15.330(043) & 14.609(031) & 14.629(035)   \\
2016 Mar. 30 & 57477.110 &  71.191 &   \nodata   & 16.178(046) & 15.027(039)  & 15.503(031) & 14.728(035) & 14.757(038)   \\
2016 Apr. 02 & 57480.830 &  74.911 & 16.855(065) & 16.252(044) & 15.045(034)  & 15.574(036) & 14.824(040) & 14.783(028)   \\
2016 Apr. 07 & 57485.535 &  79.616 & 17.352(171) & 16.305(050) & 15.126(038)  & 15.588(031) & 14.780(028) & 14.799(032)   \\
2016 Apr. 10 & 57488.785 &  82.866 & 17.925(144) & 16.512(047) & 15.213(047)  & 15.719(030) & 14.891(044) & 14.876(044)   \\
2016 Apr. 15 & 57493.775 &  87.856 & 97.478(250) & 16.829(062) & 15.361(039)  & 15.939(032) & 15.012(027) & 15.000(032)   \\
2016 Apr. 22 & 57500.465 &  94.546 &   \nodata   & 17.227(058) & 15.943(037)  & 16.401(034) & 15.444(041) & 15.363(040)   \\
2016 Apr. 26 & 57504.740 &  98.821 &   \nodata   & 17.973(062) & 16.559(040)  & 17.152(038) & 16.073(039) & 16.077(043)   \\
2016 May  01 & 57509.795 & 103.876 &   \nodata   & 18.453(085) & 16.666(047)  & 17.558(042) & 16.402(040) &   \nodata     \\
2016 May  02 & 57510.720 & 104.801 &   \nodata   & 18.464(085) & 16.815(044)  & 17.580(041) & 16.376(041) & 16.301(041)   \\
2016 May  02 & 57510.835 & 104.916 &   \nodata   & 18.445(063) & 16.891(044)  & 17.529(060) & 16.396(029) &   \nodata     \\
2016 May  04 & 57512.030 & 106.111 &   \nodata   & 18.284(093) & 16.776(038)  & 17.391(035) & 16.307(026) & 16.240(034)   \\
2016 May  04 & 57512.740 & 106.821 &   \nodata   & 18.401(059) & 16.806(040)  & 17.518(026) & 16.417(040) & 16.289(045)   \\
2016 May  05 & 57513.710 & 107.791 &   \nodata   & 18.292(065) & 16.766(049)  & 17.424(037) & 16.322(037) & 16.273(043)   \\
2016 May  10 & 57518.965 & 113.046 &   \nodata   & 18.189(057) & 16.871(055)  & 17.581(060) & 16.459(042) & 16.376(037)   \\
2016 May  11 & 57519.935 & 114.016 &   \nodata   & 18.307(069) & 16.860(033)  & 17.364(034) & 16.412(027) & 16.398(040)   \\
2016 May  12 & 57520.050 & 114.131 &   \nodata   & 18.338(045) & 16.900(041)  & 17.454(032) & 16.431(029) & 16.443(037)   \\
2016 May  14 & 57522.935 & 117.016 &   \nodata   & 18.500(112) & 16.940(034)  & \nodata     & \nodata     & \nodata       \\
2016 May  15 & 57523.123 & 117.204 &   \nodata   & 18.468(115) & 16.926(043)  & 17.485(038) & 16.460(037) & 16.388(040)    \\
2016 May  20 & 57528.760 & 122.841 &   \nodata   & 18.225(102) & 17.089(061)  &  \nodata    & 16.470(038) & 16.397(033)    \\
2016 May  28 & 57536.065 & 130.146 &   \nodata   & 18.331(035) & 17.056(037)  & \nodata     & \nodata     & \nodata       \\
2016 Jun. 04 & 57543.760 & 137.841 &   \nodata   &  \nodata    & 17.416(188)  & 17.841(029) & 16.780(032) & 16.824(024)   \\
2016 Jun. 05 & 57544.763 & 138.844 &   \nodata   & 18.729(040) & 17.372(042)  & 17.934(034) & 16.859(028) & 16.853(026)   \\
2016 Jun. 06 & 57545.885 & 139.966 &   \nodata   & 18.682(051) & 17.466(033)  & 17.927(041) & 16.883(031) & 16.851(026)   \\
2016 Jun. 07 & 57546.745 & 140.826 &   \nodata   & 18.608(066) & 17.489(038)  & \nodata     &  \nodata    &  \nodata      \\
2016 Jun. 08 & 57547.715 & 141.796 &   \nodata   & 18.734(054) & 17.494(038)  & 17.931(031) & 17.007(029) & 16.870(041)    \\
2016 Jun. 23 & 57562.705 & 156.786 &   \nodata   & 18.484(060) & 17.416(043)  & 17.855(041) & 16.812(034) &  \nodata       \\
2016 Jun. 24 & 57563.705 & 157.786 &   \nodata   & 18.612(159) & 17.516(101)  & \nodata     &  \nodata    &  \nodata      \\
2016 July 06 & 57575.745 & 169.826 &   \nodata   & 18.634(045) & 17.668(045)  & 18.069(037) & 16.871(034) & 17.084(028)    \\
 \hline
\end{tabular}
\begin{flushleft}
  $^{a}$ Relative to the explosion date, MJD = 57,405.92.\\
   \end{flushleft}
\end{table*}

\begin{table*}
\caption{UV and Optical Photometry of \sn\ from $Swift$ ($1\sigma$ Uncertainties).}  \label{tab:uvot}
 \begin{tabular}{ccccccccc}
   \hline \hline
   UT Date &    MJD       & Phase$^a$       & $uvw$2       &    $uvm$2   &    $uvw$1  &   $U$ &     $B$    & $V$   \\
   (yy/mm/dd)&            & (day)       & (mag)        &    (mag)    &    (mag)   &   (mag)&   (mag)   & (mag) \\
  \hline
  2016 Jan. 21 & 57408.07 & 2.15  & 12.74(03)  & 12.72(03) &   12.91(03)  & 13.49(03) &	14.83(04) &  14.91(06) \\
  2016 Jan. 22 & 57409.45 & 3.53  & 12.89(04)  & 12.75(04) &   12.87(04)  & 13.25(04) & 14.55(04) &  14.74(06) \\
  2016 Jan. 22 & 57409.79 & 3.87  & 12.82(04)  & 12.66(04) &   12.80(04)  & 13.20(04) & 14.49(04) &  14.57(06) \\
  2016 Jan. 23 & 57410.19 & 4.27  & 12.79(04)  & 12.62(04) &   12.73(04)  & 13.11(04) & 14.36(04) &  14.47(06) \\
  2016 Jan. 24 & 57411.82 & 5.90  & 13.00(04)  & 12.72(04) &   12.72(04)  & 12.95(04) & 14.24(04) &  14.22(05) \\
  2016 Jan. 25 & 57412.50 & 6.58  & 13.17(03)  & 12.86(03) &   12.82(03)  & 13.00(03) & 14.21(03) &  14.21(05) \\
  2016 Jan. 27 & 57414.25 & 8.33  & 13.55(04)  &  \nodata  &   \nodata    & \nodata   & \nodata   &  \nodata  \\
  2016 Feb. 01 & 57419.03 & 13.11 & 14.57(05)  &  \nodata  &   13.87(05)  & 13.28(04) & 14.29(04) &  14.13(05) \\
  2016 Feb. 01 & 57419.76 & 13.84 & 14.76(06)  &  \nodata  &   14.03(05)  & 13.36(04) & 14.30(04) &  14.18(05) \\
  2016 Feb. 06 & 57424.69 & 18.77 & 16.11(20)  &  \nodata  &   15.38(07)  & 14.05(05) & 14.49(05) &  \nodata   \\
  2016 Feb. 07 & 57425.10 & 19.18 & 16.25(08)  & 16.43(08) &   15.33(06)  & 14.06(04) & 14.53(04) &  14.28(05) \\
  2016 Feb. 08 & 57426.63 & 20.71 & 16.60(09)  & 16.83(10) &   15.67(07)  & 14.25(05) & 14.69(05) &  14.31(06) \\
  2016 Feb. 09 & 57427.96 & 22.04 & 16.92(09)  & 17.12(09) &   15.89(07)  & 14.55(05) & 14.68(05) &  14.27(05) \\
  2016 Feb. 10 & 57428.39 & 22.47 & 16.89(10)  & 17.16(11) &   15.92(08)  & 14.54(05) & 14.72(05) &  14.36(06) \\
  2016 Feb. 17 & 57435.68 & 29.76 & 18.24(19)  & 18.66(26) &   17.02(10)  & 15.59(07) &	15.14(05) &  14.48(06) \\
  2016 Feb. 21 & 57439.80 & 33.88 & 18.54(18)  & 19.39(34) &   17.21(11)  & 15.83(08) &	15.26(05) &  14.62(06) \\
  2016 Feb. 23 & 57441.59 & 35.67 & 18.40(18)  & 18.91(26) &   17.36(10)  & 16.10(08) & 15.34(05) &  14.71(07) \\
  2016 Mar. 1  & 57448.77 & 42.85 & 18.75(22)  & \nodata   &   17.64(14)  & 16.29(09) &	15.52(06) &  14.81(07) \\
  2016 Mar. 5  & 57452.46 & 46.54 & 18.95(22)  & \nodata   &   17.73(13)  & 16.39(09) & 15.61(06) &  14.80(06) \\
\hline
 \end{tabular}
\begin{flushleft}
  $^{a}$ Relative to the explosion date, MJD = 57,405.92.\\
   \end{flushleft}
\end{table*}

\begin{table*}
\caption{Observing Log for Optical Spectra of \sn.} \label{tab:spelog}
\centering
\begin{tabular}{lccccc}
\hline \hline
 UT Date	&   MJD	        &    Phase$^a$       & Range        & Exposure & Telescope + Instrument\\
            	&               &    (days)          & (\AA)        & (s)      &                       \\
\hline
2016 Jan. 20	& 57407.74	& 1.82	& 3300--9,000	& 900	& LCO 2.0~m Telescope South + FLOYDS \\
2016 Jan. 21	& 57408.52	& 2.60	& 3300--10,000	& 900	& LCO 2.0~m Telescope North + FLOYDS \\
2016 Jan. 23	& 57410.48	& 4.56	& 3250--10,000	& 900	& LCO 2.0~m Telescope North + FLOYDS \\
2016 Jan. 23	& 57410.68	& 4.76	& 3250--10,000	& 900	& LCO 2.0~m Telescope South + FLOYDS \\
2016 Jan. 23	& 57410.89	& 4.97	& 3500--9000	& 1500	& Lijiang 2.4~m + YFOSC \\
2016 Jan. 25	& 57412.47	& 6.55	& 3250--10,000	& 900	& LCO 2.0~m Telescope North + FLOYDS \\
2016 Jan. 25	& 57412.67	& 6.75	& 3250--10,000	& 900	& LCO 2.0~m Telescope South + FLOYDS \\
2016 Jan. 27	& 57414.52	& 8.60	& 3250--10,000	& 900	& LCO 2.0~m Telescope North + FLOYDS \\
2016 Jan. 28	& 57415.54	& 9.62	& 3250--10,000	& 1200	& LCO 2.0~m Telescope North + FLOYDS \\
2016 Jan. 28	& 57415.93	& 10.01	& 3500--9100	& 1200	& Lijiang 2.4~m + YFOSC \\
2016 Jan. 31	& 57418.73	& 12.81	& 3350--10,000	& 1200	& LCO 2.0~m Telescope South + FLOYDS \\
2016 Feb. 2	& 57420.88	& 14.96 & 3500--9100	& 1800	& Lijiang 2.4~m + YFOSC \\
2016 Feb. 3	& 57421.46	& 15.54	& 3250--10,000	& 1200	& LCO 2.0~m Telescope North + FLOYDS \\
2016 Feb. 6	& 57424.71	& 18.79	& 3400--10,000	& 1200	& LCO 2.0~m Telescope South + FLOYDS \\
2016 Feb. 10	& 57428.68	& 22.76	& 3400--10,000	& 1200	& LCO 2.0~m Telescope South + FLOYDS \\
2016 Feb. 13	& 57431.90	& 25.98 & 3500--9100	& 1500	& Lijiang 2.4~m + YFOSC \\
2016 Feb. 16	& 57434.69	& 28.77	& 3550--10,000	& 1200	& LCO 2.0~m Telescope South + FLOYDS \\
2016 Feb. 16	& 57434.86	& 28.94 & 3500--9100	& 1500	& Lijiang 2.4~m + YFOSC \\
2016 Feb. 18	& 57436.87	& 30.95	& 3500--9100	& 1500	& Lijiang 2.4~m + YFOSC \\
2016 Feb. 22	& 57440.47	& 34.55	& 3550--10,000	& 1200	& LCO 2.0~m Telescope North + FLOYDS \\
2016 Feb. 23	& 57441.84	& 35.92	& 3500--9100	& 1500	& Lijiang 2.4~m + YFOSC \\
2016 Feb. 28	& 57446.54	& 40.62	& 3500--10,000	& 1200	& LCO 2.0~m Telescope North + FLOYDS \\
2016 Feb. 28	& 57446.84	& 40.92 & 3500--9100	& 1500	& Lijiang 2.4~m + YFOSC \\
2016 Mar. 6	& 57453.64	& 47.72	& 3500--10,000	& 1200	& LCO 2.0~m Telescope North + FLOYDS \\
2016 Mar. 11	& 57458.86	& 52.94 & 3500--9100	& 1800	& Lijiang 2.4~m + YFOSC \\
2016 Mar. 12	& 57459.69	& 53.77	& 3700--10,000	& 1200	& LCO 2.0~m Telescope South + FLOYDS \\
2016 Mar. 18	& 57465.51	& 59.59	& 3600--10,000	& 1200	& LCO 2.0~m Telescope North + FLOYDS \\
2016 Mar. 18	& 57465.75	& 59.83 & 3500--9100	& 1800	& Lijiang 2.4~m + YFOSC \\
2016 Mar. 21	& 57468.74	& 62.82	& 3950--10,000	& 1200	& LCO 2.0~m Telescope South + FLOYDS \\
2016 Mar. 27	& 57474.71	& 68.79	& 3700--9150	& 2850	& Lijiang 2.4~m + YFOSC \\
2016 Apr. 4	& 57482.72	& 76.80	& 3650--9150	& 2100	& Lijiang 2.4~m + YFOSC \\
2016 Apr. 7	& 57485.51	& 79.59	& 3800--10,000	& 1200	& LCO 2.0~m Telescope South + FLOYDS \\
2016 Apr. 13	& 57491.69	& 85.77	& 3900--10,000	& 1200	& LCO 2.0~m Telescope South + FLOYDS \\
2016 Apr. 17	& 57495.69	& 89.77	& 3900--8780	& 2100	& Xinglong 2.16~m + BFOSC \\
2016 Apr. 17	& 57495.80	& 89.88	& 3600--9100	& 2100	& Lijiang 2.4~m + YFOSC \\
2016 Apr. 26	& 57504.72	& 98.80	& 3600--9100	& 2100	& Lijiang 2.4~m + YFOSC \\
2016 Apr. 29	& 57507.26	& 101.34& 3500--10,000	& 1200	& LCO 2.0~m Telescope North + FLOYDS \\
2016 May 2	& 57510.72	& 104.80& 3500--9170	& 2100	& Lijiang 2.4~m + YFOSC \\
2016 May 21	& 57529.27	& 123.35& 4500--9300	& 1800	& LCO 2.0~m Telescope North + FLOYDS \\
2016 Jun. 9	& 57548.34	& 142.42& 3500--10,000	& 3600	& LCO 2.0~m Telescope North + FLOYDS \\
   \hline
 \end{tabular}
\begin{flushleft}
  $^{a}$ Relative to the explosion date, MJD = 57,405.92.
   \end{flushleft}
\end{table*}

\begin{table*}
\caption{Photometric Parameters of \sn.} \label{tab:photpara}
\begin{tabular}{ccccccccc}
\hline \hline
 &  $U$ &  $B$  & $g$  & $V$   & $r$   & $R$   &  $i$ & $I$ \\
 &  (mag) & (mag) & (mag) & (mag) & (mag) & (mag) & (mag) & (mag)  \\
\hline
Peak magnitude 		&    13.25 & 14.14 & 14.04 & 14.05  & 13.99 & 13.91 & 14.07 & 13.77 \\
Phase of maximum$^{a}$ &   9.26  &  9.60 & 10.60 & 11.26  & 13.70 & 13.76 & 13.55 & 14.11 \\
Plateau magnitude	&    --    &  --   & 15.31 & 14.67  & 14.60 & 14.46 & 14.50 & 14.07 \\
Decay rate (mag/100 d)	&   --     & 0.58  & 0.99  & 1.35   & 1.25  & 1.22  & 1.05  & 1.14  \\
\hline
\end{tabular}
\begin{flushleft}
  $^{a}$ Relative to the explosion date, MJD = 57,405.92.\\
   \end{flushleft}
\end{table*}

\begin{table*}
\centering
\caption{The mass and radius of SNe IIP using direct archive images and hydrodynamic model method.} \label{tab:massnrad}
\begin{tabular}{|c|c|c|c|c|c|}
\hline \hline
\multirow{2}{*}{SN name} &
\multicolumn{2}{c|}{HST image} &
\multicolumn{2}{|c}{Modeling} &
\multirow{1}{*}{References}\\
& mass  & radius  & mass   & radius  \\
& (\msun) & (\rsun) & (\msun)  & (\rsun) \\
\hline  
SN 2005cs & $9_{-2}^{+3}$  &  \nodata & 11 & 360 $\pm$ 70 & \cite{2005MNRAS.364L..33M, 2017MNRAS.464.3013P}\\
SN 2008bk & $12.9_{-1.8}^{+1.6}$ & 470 $\pm$ 16& 12 & 502 &   \cite{2014MNRAS.438.1577M, 2017MNRAS.466...34L}\\
SN 2009N & \nodata  &  \nodata    & 13 $\pm$ 2 & 287$\pm$ 43 & \cite{2014MNRAS.438..368T}\\
SN 2009md & $8.5_{-1.5}^{+6.5}$ & \nodata & 10 & 288 & \cite{2011MNRAS.417.1417F, 2017MNRAS.464.3013P} \\
SN 2012A & $10.5_{-2}^{+4.5}$  &    \nodata  & 14 $\pm$ 2 & 260 $\pm$ 40 & \cite{2013MNRAS.434.1636T} \\
SN 2012aw & 14 -- 26 & 1040 $\pm$ 100 & 22 -- 24 & 290 -- 580 & \cite{2012ApJ...756..131V, 2012ApJ...759L..13F, 2014ApJ...787..139D}  \\
SN 2012ec & 14 -- 22 &1030 $\pm$ 180 & 14.0 -- 14.6   & 230 $\pm$ 70 & \cite{2013MNRAS.431L.102M, 2015MNRAS.448.2312B}  \\
SN 2013ej & 8 -- 15.6  & \nodata     & 12.5 $\pm$ 1.9 &415 $\pm$ 62& \cite{2014MNRAS.439L..56F, 2015ApJ...807...59H}  \\
SN 2016X  &  \nodata   &  \nodata& 18.5 -- 19.7 &  925 $\pm$ 65 & this  work \\
\hline
\end{tabular}
\end{table*}


\bsp	
\label{lastpage}
\end{document}